\documentclass[jcp,preprint]{revtex4-1}

\usepackage{graphicx}
\usepackage{amstext}
\usepackage{amsmath}
\usepackage{amssymb} 
\usepackage{color}
\usepackage{braket}
\usepackage[acronym]{glossaries} %

\def\be{\begin{equation}}
\def\ee{\end{equation}}
\def\bea{\begin{eqnarray}}
\def\eea{\end{eqnarray}}
\def\bal#1\eal{\begin{align*}#1\end{align*}}
\def\ba#1\ea{\begin{align}#1\end{align}}
\def\be{\begin{eqnarray}}
\def\ee{\end{eqnarray}}

\newcommand{\fig}[1]{Fig.~\ref{#1}}

\newcommand{\eq}[1]{Eq.~\eqref{#1}}

\newcommand{\eqs}[1]{Eqs.~\eqref{#1}}

\date{\today}

\newacronym{CI}{CI}{conical intersection} %
\newacronym{GP}{GP}{geometric phase} %
\newacronym{LVC}{LVC}{linear vibronic coupling} %
\newacronym{DOF}{DOF}{degrees of freedom} %
\newacronym{NAC}{NAC}{nonadiabatic coupling} %
\newacronym{DBOC}{DBOC}{diagonal Born-Oppenheimer correction} %
\begin{document}

\title{Geometric phase effects in excited state dynamics through a conical intersection in large molecules:
N-dimensional linear vibronic coupling model study}

\author{Jiaru Li}
\affiliation{Department of Physical and Environmental Sciences,
  University of Toronto Scarborough, Toronto, Ontario, M1C 1A4,
  Canada}
  
\author{Lo{\"i}c Joubert-Doriol}
\affiliation{Department of Physical and Environmental Sciences,
  University of Toronto Scarborough, Toronto, Ontario, M1C 1A4,
  Canada}
\affiliation{Chemical Physics Theory Group, 
  Department of Chemistry, University of Toronto, Ontario, M5S 3H6, 
  Canada}
  
\author{Artur F. Izmaylov}
\email{artur.izmaylov@utoronto.ca}
\affiliation{Department of Physical and Environmental Sciences,
  University of Toronto Scarborough, Toronto, Ontario, M1C 1A4,
  Canada}
\affiliation{Chemical Physics Theory Group, 
  Department of Chemistry, University of Toronto, Ontario, M5S 3H6, 
  Canada}
  
\begin{abstract}
We investigate geometric phase (GP) effects 
in nonadiabatic transitions through a conical intersection (CI)  
in an N-dimensional linear vibronic coupling (ND-LVC) model. 
This model allows for the coordinate transformation encompassing all nonadiabatic effects within 
a two-dimensional (2D) subsystem while the other N-2 dimensions form a system of uncoupled 
harmonic oscillators identical for both electronic states and coupled bi-linearly with the 
subsystem coordinates. The 2D subsystem governs ultra-fast nonadiabatic dynamics through the CI 
and provides a convenient model for studying GP effects.
Parameters of the original ND-LVC model define the Hamiltonian of the transformed 
2D subsystem and thus influence GP effects directly. Our analysis reveals what values of ND-LVC 
parameters can introduce symmetry breaking in the 2D subsystem that diminishes GP effects.   
\end{abstract}

\maketitle

\section{Introduction}

\Glspl{CI} of potential energy surfaces are one of the most frequent reasons for break-down of the Born-Oppenheimer approximation in molecules beyond diatomics.\cite{Truhlar:2003/pra/032501,Migani:2004/271,Domcke:2012/arpc/325,Yarkony:1996/rmp/985} 
Besides promoting nonadiabatic transitions between involved electronic 
states, due to their non-trivial topology, \glspl{CI} also give rise to \glspl{GP} in the adiabatic electronic and nuclear wavefunctions.\cite{LonguetHigg:1958/rspa/1,Berry:1984/rspa/45,Mead:1979/jcp/2284, Berry:1987/rspa/31,Schon:1995/jcp/9292,JuanesMarcos:2005/sci/1227,Kendrick:2003,Hazra:2015he} The \gls{GP} of the electronic wavefunction results in the wavefunction sign change upon continuous parametric evolution around the \gls{CI} in the nuclear configuration space.\cite{LonguetHigg:1958/rspa/1,Berry:1984/rspa/45}
This sign change makes the electronic wavefunction double-valued. To preserve the single-valued character of the total electron-nuclear wavefunction, the nuclear counterpart must also be double-valued. 

It has been shown that failure to account for the \gls{GP} can lead to substantial deviations from the exact dynamics in symmetric molecules (e.g., butatriene cation, pyrazine, and phenol).\cite{Althorpe:2008/jcp/214117,Joubert:2013/jcp/234103,Ryabinkin:2014/jcp/214116,Bouakline:2014/cp/31,Xie:2016/jacs/7828} This is relevant for nuclear dynamics near \glspl{CI} independent of whether the nuclear wave-packet is on a lower or higher electronic potential energy surface. Thus, including more than one electronic state does not free from the necessity to account for the \gls{GP}.\cite{Althorpe:2008/jcp/214117,Ryabinkin:2014/jcp/214116,Bouakline:2014/cp/31}   

To consider the importance of \gls{GP} effects in large molecules such as photo-active proteins (e.g., rhodopsin), one should address growing number of nuclear \gls{DOF}.\cite{Gozem:2012kg,Tscherbul:pccp/2015} On the one hand, it is well-known that quantum effects based on a wave nature of quantum particles can be diminished with increasing the system size. If the number of involved \gls{DOF} is growing one generally arrives to the 
classical limit. On the other hand, the previous consideration of \gls{GP} effects in low-energy dynamics shown that extra nuclear \gls{DOF} may not always reduce the importance of the \gls{GP}. \cite{Joubert:2013/jcp/234103} In this paper we analyze how the importance of GP effects can be affected by a large collection of nuclear DOF when nuclear dynamics is initiated on the excited electronic state.
  
To model a large number of nuclear DOF participating in the nonadiabatic dynamics through a CI, 
we consider a general 2-state $N$-dimensional \gls{LVC} model\cite{Koppel:1984/acp/59} 
whose Hamiltonian in the diabatic representation is
\begin{equation}\label{eq:HND}
H_{\rm ND} = \sum_{j}^{N}\frac{1}{2}(p_j^2 + \Omega_j^2q_j^2)\boldsymbol{I}_2 + \begin{bmatrix} \tilde{\kappa}_jq_j & \lambda_jq_j \\ \lambda_jq_j & \kappa_jq_j \end{bmatrix} + \begin{bmatrix} -\delta/2 & 0 \\ 0 & \delta/2 \end{bmatrix},
\end{equation}
where $p_j$ and $q_j$ are momentum and position of the $j^{th}$ coordinate, $\Omega_j$ are frequencies, $\boldsymbol{I}_2$ is the identity 2-by-2 matrix, $\kappa_j$, $\tilde\kappa_j$ and $\lambda_j$ are linear couplings of electronic states, and $\delta$ is the energy gap between the two electronic states at the origin.
Atomic units are used throughout the paper.
One can consider the ND-LVC model as the electronic two-state system embedded in the environment of nuclear DOF. Although for studying GP effects we will need the adiabatic representation, our starting point is the diabatic Hamiltonian because the diabatic-to-adiabatic transformation is uniquely defined while this is not generally true for the inverse transformation.\cite{Mead:1982/jcp/6090}    
In spite of its simplicity, the ND-LVC Hamiltonian has a wide range of applications \cite{Cederbaum:1977/cp/169,Sukharev:2005/pra/012509,Gindensperger:2006/jcp/144104,Izmaylov:2011/jcp/234106} and can be further utilized to model the \gls{CI} vicinity of more complex topographies.\cite{Koppel:1984/acp/59,Thiel:1999/jcp/9372,Koppel:2001/jcp/2377} However, in practice, when $N$ is large, performing ND system simulations becomes computationally expensive. The complexity of dynamical simulations becomes even higher in the adiabatic representation due to divergence of the nonadiabatic couplings at the CI seam. 
Previous studies~\cite{Cederbaum:2005/prl/113003,Gindensperger:2006/jcp/144103,Gindensperger:2006/jcp/144104,Ryabinkin:2014/jcp/214116} suggest that there exists a transformation of the ND Hamiltonian to a Hamiltonian where nonadiabatic effects are confined within a two-dimensional subspace of effective nuclear variables coupled to the $N-2$ other nuclear DOF in an identical way for two electronic states. The 2D subsystem of the transformed Hamiltonian can be used to simulate short-time nonadiabatic dynamics, and its parameters are functions of those from the original ND-LVC model.\cite{Ryabinkin:2014/jcp/214116} Therefore, we will explore the influence of an increasing number of nuclear DOF in the ND-LVC model on GP effects through considering GP effects in the corresponding 2D subsystem model.

The rest of this paper is organized as follows. Section \ref{sec:theory} introduces the transformation from the ND-LVC model 
to the 2D-LVC subsystem immersed into an $N-2$ dimensional bath, 
explores the geometry of the 2D-LVC model, and discusses the definition and main manifestations of GP effects.  
Section \ref{sec:res_disc} provides numerical examples of illustrative 2D subsystem parameter sets 
and corresponding nonadiabatic dynamics. Finally, Section \ref{sec:conc} concludes by summarizing 
main findings.

\section{Theory}
\label{sec:theory}


\subsection{Model for the effective two-dimensional Hamiltonian}

Following steps presented in Appendix~\ref{app:transfo}, the $N$-dimensional Hamiltonian in \eq{eq:HND} can 
be transformed into a sum 
\bea\label{eq:Htot}
H = H_S + H_{SB} + H_B.
\eea
Here
\bea\label{eq:H2D}
H_S &=& T_S\boldsymbol{I}_2 + \begin{bmatrix}V_{11}(x,y) & V_{12}(x,y) \\ V_{12}(x,y) & V_{22}(x,y) \end{bmatrix},
\eea
is the subsystem Hamiltonian encompassing all nonadiabatic effects of the full problem 
within a 2D subspace of collective nuclear 
variables $x$ and $y$, which are linear combinations of $\{q_j\}$.
The first term in $H_S$ is the nuclear kinetic energy operator, $T_S = -1/2 (\partial^2/\partial x^2 + \partial^2/\partial y^2)$, 
multiplied by the 2 by 2 electronic identity matrix $\boldsymbol{I}_2$. 
The second term in $H_S$ is the diabatic potential matrix with two harmonic potentials
\bea
    V_{11}(x,y) = \frac{1}{2}[\omega_x^2(x-x_0)^2+\omega_y^2(y-y_0)^2-\Delta] \\
    V_{22}(x,y) = \frac{1}{2}[\omega_x^2(x+x_0)^2+\omega_y^2(y+y_0)^2+\Delta]
\eea
and linear coupling 
\begin{equation}
    V_{12}(x,y) = c_xx + c_yy + \Delta_{12}.
\end{equation}
$H_B$ in \eq{eq:Htot} represents the $(N-2)$-dimensional bath Hamiltonian of shifted uncoupled harmonic oscillators
\bea
H_B &=& \frac{1}{2} \sum_{j=1}^{N-2} (P_j^2+\tilde\Omega_j^2 Q_j^2+f_jQ_j) \boldsymbol{I}_2,
\eea
with collective bath coordinates and momenta, $Q_j$ and $P_j$, respectively. Note that parameters for 
all bath oscillators are identical for both electronic states.
Finally, $H_{SB}$ of \eq{eq:Htot} describes bi-linear coupling between the subsystem and bath coordinates
 and momenta
 \bea
   H_{SB} &=& \sum_{j=1}^{N-2} \Bigg[\Lambda_{jx}\Bigg( \sqrt{\omega_x\tilde\Omega_j} x Q_j + \frac{ p_x P_j}{\sqrt{\omega_x\tilde\Omega_j}} \Bigg) \nonumber\\
           & &\hspace{1cm}+\Lambda_{jy}\Bigg( \sqrt{\omega_y\tilde\Omega_j} y Q_j + \frac{ p_y P_j}{\sqrt{\omega_y\tilde\Omega_j}} \Bigg)\Bigg] \boldsymbol{I}_2,
\eea 
these nuclear couplings are identical for both diabatic electronic states.
All parameters in $H_S$, $H_{SB}$, and $H_B$ are functions of those of $H_{\rm ND}$ (\eq{eq:HND})
and are defined in Appendix~\ref{app:transfo}.

This paper focuses on processes where the system starts in the minimum of diabatic electronic state one and 
is instantaneously photo-excited to the higher electronic state. Hence, the initial wave-packet is centred at the 
Franck-Condon point that is at the minimum of state one. For the \gls{LVC} model with the minimum 
of the first state in the origin of the coordinate system, we have $\sum_j\tilde{\kappa}_j^2=0$ in \eq{eq:HND}. 
For an initial wave-packet that is not centred at the minimum of one of the two diabats, a third coordinate 
must be included in the subsystem Hamiltonian. This case will not be considered here.

We will consider GP effects in the 2D subsystem instead of the original ND system, 
assuming that the parameters responsible for the subsystem-bath interaction, $\Lambda_{ix}$ and $\Lambda_{iy}$, are smaller than the nonadiabatic couplings. This is equivalent to the assumption that the time-scale of the subsystem-bath interaction is much slower than that of the subsystem dynamics. The strength and the effect of the subsystem-bath couplings are discussed in Appendix~\ref{app:transfo}. Our previous work confirmed weakness of the system-bath couplings for a series of typical benchmark systems 
like butatriene cation and pyrazine.\cite{Ryabinkin:2014/jcp/214116} 
In this setup effects of nuclear DOF on dynamics near a CI is presented through the $H_S$ parameters' 
dependence on those of \eq{eq:HND}.
Variations of the model Hamiltonian (e.g. inclusion of higher order terms) would make the transformation from \eq{eq:HND} to \eq{eq:Htot} (see Appendix~\ref{app:transfo}) exact only at a given geometry and therefore valid in a restricted region of the nuclear space. In this case, neglecting bath nuclear \gls{DOF} can lead to more 
significant deviations from the exact dynamics.

\begin{figure}[ht]
  \centering
    \includegraphics[width=0.5\textwidth]{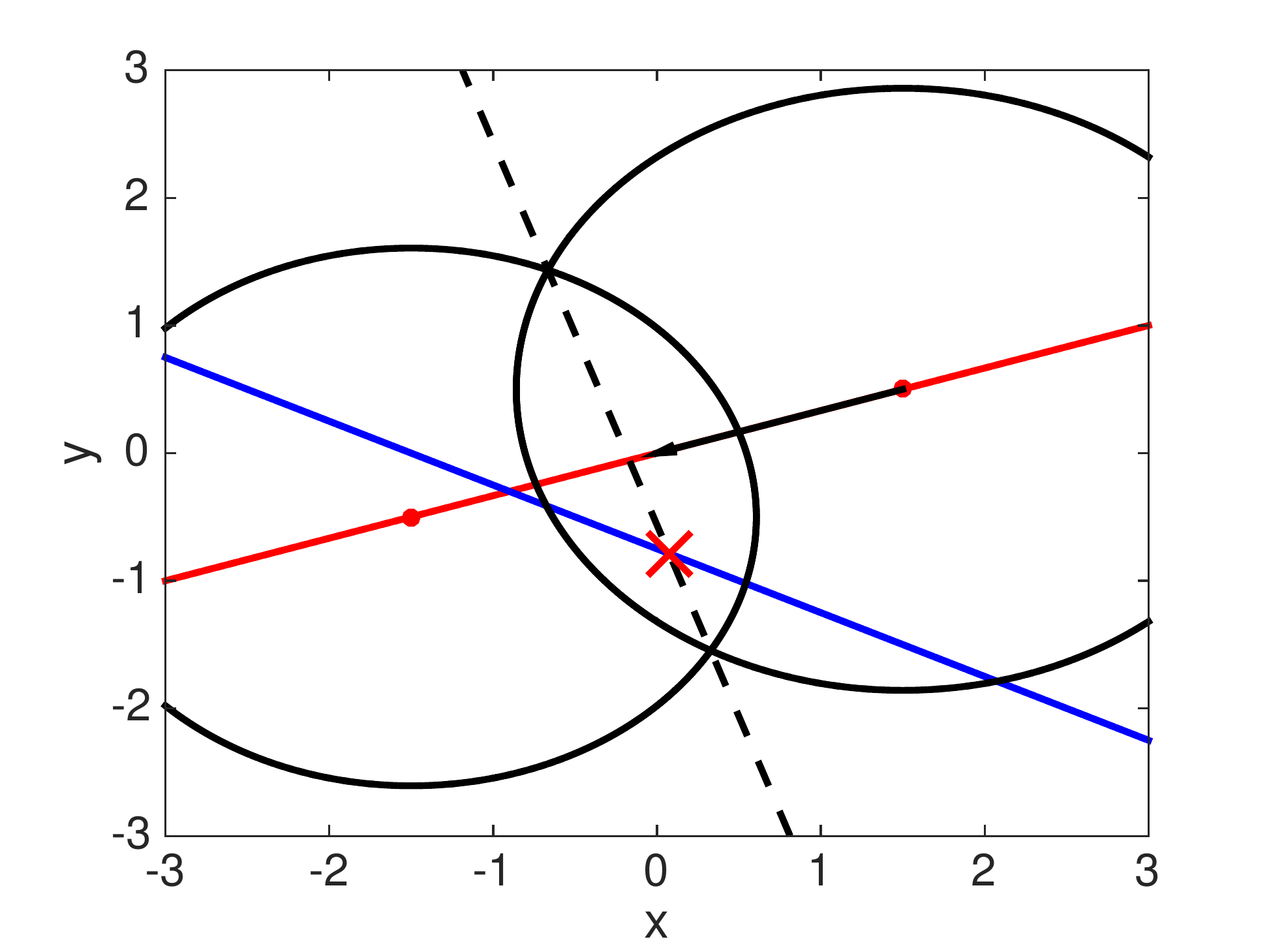}
  \caption{Top view of a general 2D subsystem:  
  the tuning direction (solid red), the zero-coupling line (solid blue),
  and the degeneracy line (dashed black). The solid black ellipsoids are level lines for $V_{11}$ and $V_{22}$ potentials. The red cross indicates the position of the CI.}
\label{fig:paramgeom}
\end{figure}

The parameters of $H_S$ have geometrical meaning illustrated in \fig{fig:paramgeom} and summarized as follows:
\begin{itemize}
\item Vector $(2x_0,2y_0)$ gives the relative positions of diabatic potential minima and defines the tuning direction.
\item Vector $(\omega_x^2x_0,\omega_y^2y_0)$ is normal to the degeneracy line, $V_{11}=V_{22}$.
\item Vector $(c_x,c_y)$ defines the coupling direction, which is normal to the zero coupling line, $V_{12}=0$.
\item Parameter $\Delta$ is the energy difference between diabatic potential minima.
\item Parameter $\Delta_{12}$ determines the displacement of the zero coupling line along the coupling direction.
\end{itemize}

To study \gls{GP} effects in a general 2D model, we define a symmetric Hamiltonian model with $y_0=c_x=\Delta=\Delta_{12}=0$.
To explore different cases in \eq{eq:H2D} we will add variations $\Delta_{12}\ne 0$, $\Delta\ne 0$, and $c_x\ne 0$ to this symmetric 
setup. For the symmetric model both diabatic potentials have the same energies and the zero coupling line coincides with the tuning direction and is orthogonal to the degeneracy line (see \fig{fig:paramgeom}). Variations change geometry of the model and as a result the CI position 
\bea\label{eq:ci_coor}
(x_{\rm CI},y_{\rm CI}) & = & \left(-\frac{\Delta}{2\omega_x^2x_0},-\frac{\Delta_{12}}{c_y}+\frac{c_x\Delta}{2c_y\omega_x^2x_0}\right)
\eea
because the CI is always located on the intersection of the degeneracy and zero-coupling lines (\fig{fig:paramgeom}).

\subsection{Definition of geometric phase effects} 

To analyze GP effects one needs to consider the adiabatic representation, which can be obtained by diagonalizing the potential matrix in \eq{eq:H2D} 
\bea
H_{\text{adi}}&=& U^{\dag} H_S U \\
\label{eq:Hadi}
&=& \begin{bmatrix}T_S + \tau_{11} & \tau_{12} \\ \tau_{21} & T_S+ \tau_{22}\end{bmatrix} 
+\begin{bmatrix}W_- & 0 \\ 0 & W_+\end{bmatrix}. 
\eea
Adiabatic potential energy surfaces ($W_{\pm}$) are the eigenvalue of the diabatic potential matrix
\begin{equation}
        W_{\pm} = \frac{1}{2}(V_{11} + V_{22}) \pm \frac{1}{2} \sqrt{(V_{11}-V_{22})^2+4V_{12}^2}.
\end{equation}
The diagonalization requires a unitary rotation matrix 
\begin{equation}\label{eq:D2Arotation}
         U  =         
        \begin{bmatrix}
        \cos\theta  &  \sin\theta \\
        -\sin\theta &  \cos\theta \\
        \end{bmatrix}
\end{equation}
where 
\begin{equation}\label{eq:mixangle}
        \theta = \frac{1}{2}\arctan\left(\frac{2V_{12}}{V_{22}-V_{11}}\right). 
\end{equation}
Since the $U$ rotation  depends on the nuclear position, the \glspl{NAC} $\tau_{ij}$ appear in \eq{eq:Hadi} as a result of non-commutativity between the kinetic energy and the $U$ rotation. Term $\tau_{ii}$ is known as the \gls{DBOC}, which acts as a repulsive potential\cite{Ryabinkin:2014/jcp/214116,Gherib:2016ch}
\begin{equation}\label{eq:DBOC}
    \tau_{ii}  = \frac{1}{2}\nabla\theta \cdot \nabla\theta.
\end{equation}
The off-diagonal \glspl{NAC} 
\begin{equation}\label{eq:NAC}
    \tau_{12} = -\tau_{21} = -\frac{1}{2}\nabla^2\theta - \nabla\theta \cdot \nabla
\end{equation}
enable nonadiabatic transitions in the adiabatic representation.
 
The diabatic and adiabatic representations are equivalent since there is the unitary transformation, $U$, which connects corresponding Hamiltonians. However, there is a complication in the adiabatic representation associated with 
double-valued boundary conditions (BC) of electronic and nuclear wavefunctions. 
The electronic functions of the adiabatic representation can be expressed as 
\bea
\ket{\phi_1}  &=& \cos\theta\ket{1} - \sin\theta\ket{2} \\
\ket{\phi_2}  &=& \sin\theta\ket{1} + \cos\theta\ket{2}
\eea
where $\ket{1}$ and $\ket{2}$ are the diabatic electronic states. Angle $\theta$ changes by $\pi$ if one considers 
a continuous evolution of $\{\ket{\phi_i}\}_{i=1,2}$ along any closed contour around the CI, 
which makes $\{\ket{\phi_i}\}_{i=1,2}$ double-valued functions of nuclear coordinates $x$ and $y$. This is purely geometric effect associated with the presence of the CI.
The sign change can be presented as a phase factor $e^{i\theta}$, where 
$\theta$ defines the geometric phase \cite{LonguetHigg:1958/rspa/1,Berry:1984/rspa/45}.
In order to preserve a single-valued character of the total electron-nuclear wavefunction in the adiabatic representation, corresponding nuclear wavefunctions should be also obtained using double-valued BC. Imposing such BC complicates simulations because usual nuclear basis functions (e.g., gaussians) are single-valued. An alternative treatment that accounts for the GP and avoids using double-valued functions was introduced by Mead and Truhlar\cite{Mead:1979/jcp/2284}, it uses a gauge freedom in definitions of electronic and nuclear wavefunctions of the adiabatic representation, in other words, one can always consider complex but single-valued electronic eigenfunctions obtained as $\ket{\tilde{\phi}_j} = e^{i\theta}\ket{\phi_j}$. Formally this is equivalent to considering a different nuclear Hamiltonian for the adiabatic representation 
\bea
\label{eq:H_GP1}
H_{\rm GP} &=& e^{-i\theta}H_{\text{adi}}e^{i\theta} \\
\label{eq:H_GP}
&=& \begin{bmatrix}T_S + \tau_{11}^{\rm (GP)} & \tau_{12}^{\rm (GP)} \\ 
\tau_{21}^{\rm (GP)} & T_S+ \tau_{22}^{\rm (GP)}\end{bmatrix} 
+\begin{bmatrix}W_- & 0 \\ 0 & W_+\end{bmatrix}. 
\eea
Nuclear dynamics with $H_{\rm GP}$ is equivalent to that with the diabatic Hamiltonian $H_S$ for the single-valued nuclear wavefunctions. Thus, to study GP effects we contrast results of $H_{\rm adi}$ and $H_{\rm GP}$ with the same single-valued BC. The phase factors in \eq{eq:H_GP1} produce extra terms as a result of the action of the nuclear kinetic energy operator
\bea\label{eq:NAC_GP}
\tau_{12}^{\text{(GP)}} & =  & -2\tau_{11} + \tau_{12} \\
            \label{eq:DBOC_GP}
\tau_{jj}^{\text{(GP)}} &=& i\tau_{12} + 2\tau_{jj},
\eea
where $\tau_{ij}$ are given by \eqs{eq:DBOC} and \eqref{eq:NAC}. 

\subsection{GP effects in 2D models}

\begin{figure*}[ht]
  \centering
    \includegraphics[width=1\textwidth]{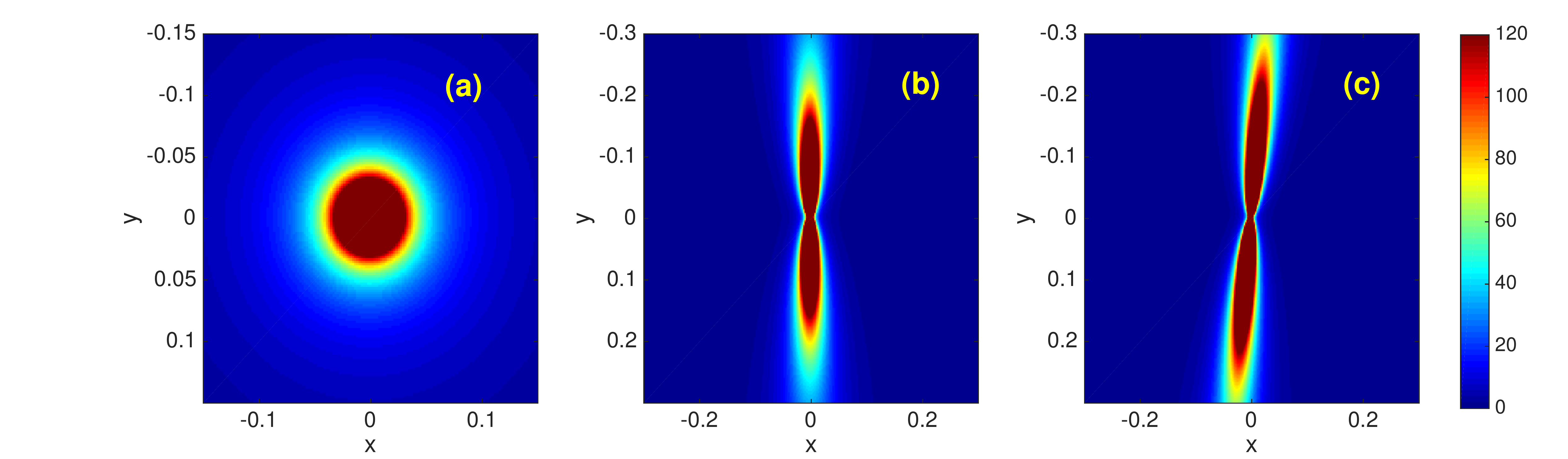}
  \caption{DBOC for different values of $\gamma$ and directions of coupling: (a) $\gamma = 1$ and $\beta = 0$, (b) $\gamma = 0.2$ and $\beta=0$, (c) $\gamma = 0.2$ and $\beta = 3$}.
\label{fig:DBOCshape}
\end{figure*}

\paragraph{DBOC compensation:} 
Without GP the repulsive \gls{DBOC} term given by \eq{eq:DBOC} can prevent a nuclear wave-packet to approach regions of strong nonadiabatic coupling (large $\tau_{12}$). This effect is most important when the kinetic energy of the wave-packet is low. Adding the GP introduces extra terms  in \eqs{eq:DBOC_GP} and \eqref{eq:NAC_GP} so that the overall repulsive effect of the DBOC is compensated.\cite{Ryabinkin:2014/jcp/214116} Thus the importance of this compensating GP effect is directly related to the significance of the DBOC in the nonadiabatic dynamics without GP. For a general 2D system given by $H_S$ [\eq{eq:H2D}] the DBOC is given by
\bea\label{eq:DBOC_2d}
    \tau_{ii} = \frac{\Delta y^2+\Delta x^2}{8[\gamma^{-1}(\Delta x)^2+
    \gamma(\Delta y+\beta \Delta x)^2]^2},
\eea
where $\Delta y = y-y_{\rm CI}$ and $\Delta x = x-x_{\rm CI}$ are distances from the CI, $\gamma = c_y/(w_x^2x_0)$ is the coupling strength, and $\beta = c_x / c_y$ is the tilting slope between the coupling and tuning directions. 
The DBOC diverges at the CI $(x_{\rm CI},y_{\rm CI})$, but what is more important is its rate of growth in different directions. DBOC's growth anisotropy is regulated by $\gamma$, for $\gamma=1$ the DBOC is cylindrical while $\gamma\ne 1$ produces anisotropic DBOC (\fig{fig:DBOCshape}). 
The impact of the DBOC on nonadiabatic dynamics depends on how large the DBOC is in the region of space accessible to a nuclear wavepacket. In a common scenario of $x$-coordinate being the tuning direction ($\beta=0$ and $y_0=0$) the DBOC importance will depend on how extended it is in the $y$ direction and how far the CI point from the origin of the coordinate system. The DBOC extension in the $y$ direction is growing with $\gamma^{-1}$ (\fig{fig:DBOCshape}).

\paragraph{Transition enhancement:}
The second GP effect is enhancement of nonadiabatic transfer for certain components of a nuclear wave-packet. This can be easily seen by considering the $\tau_{12}$ term governing nonadiabatic transfer for the special symmetric case of $\gamma=1$ and $\beta=0$ 
\bea
\tau_{12} &=& -\frac{1}{2r^2} \frac{\partial}{\partial \phi} = -\frac{i}{2r^2}L_z.
\eea 
Here, $\tau_{12}$ is written in the polar coordinates centred at the CI with radius $r$ and angle $\phi$. Any nuclear wavefunction, $\chi(x,y,t)$, can also be written as a linear combination of $L_z$ eigenfunctions centred at the CI, 
with coefficients that are dependent on the radius and time
\begin{equation}
\chi(x,y,t) = \sum_{m=-\infty}^{\infty}C_m(r,t)e^{-im\phi}.
\end{equation}
The efficiency of the nonadiabatic transition can be estimated by applying $\tau_{12}$ on $\chi(x,y,t)$. One special term in the summation is the $m=0$ term. Since $L_ze^{-im\phi}\vert_{m=0} = 0$, the action of $\tau_{12}$ on this term is 0, which implies no transfer for the $m=0$ component.

Once the GP is included, the transfer of the $m=0$ component becomes possible
\bea
\tau_{12}^{\text{(GP)}}  C_0(r,t) =  \frac{C_0(r,t)}{8r^2},  
\eea
because  $\tau_{12}^{\text{(GP)}}$ has an additional component arising from the GP
\begin{equation}\label{eq:NAC_GP_2D}
\tau_{12}^{\text{(GP)}} =  \frac{-iL_z+1/2}{2r^2}.
\end{equation}
Thus, a significant difference between dynamics with and without GP should be expected when a wave-packet  has a large portion of the $m=0$ component when it arrives at the vicinity of the CI.\cite{Ryabinkin:2014/jcp/214116}

\begin{figure*}[ht]
  \centering
    \includegraphics[width=1\textwidth]{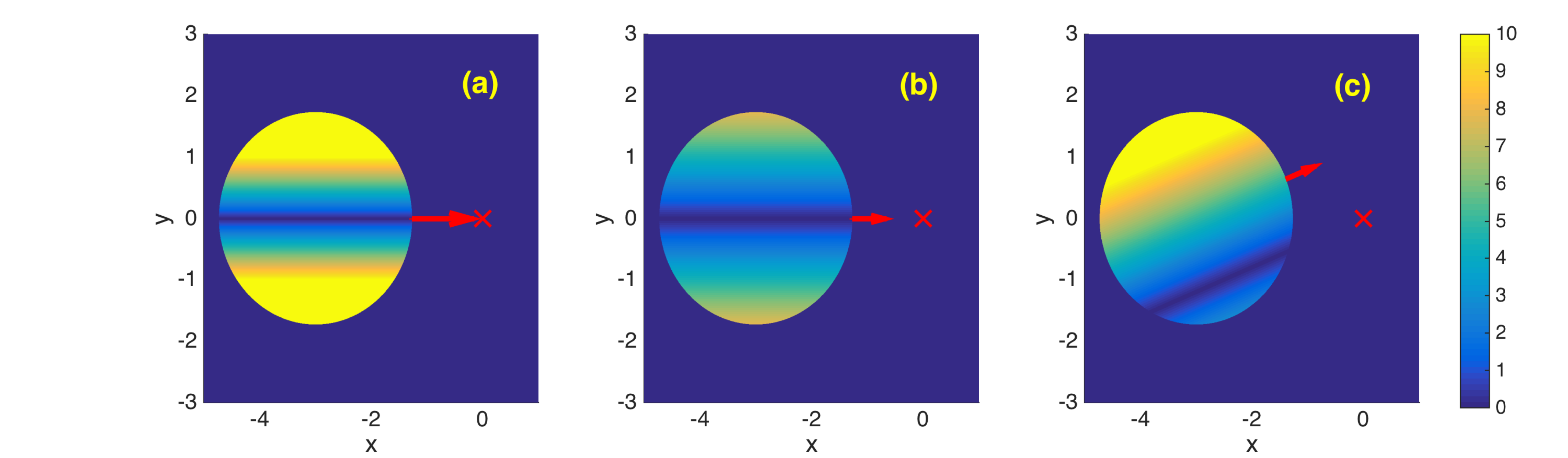}
  \caption{Circles represent nuclear wave-packets approaching CIs (red crosses). 
  The colormap shows the absolute value of the classical angular momentum for each point 
  of the wave-packets: (a) high head-on velocity of 10 a.u., (b) low head-on velocity of 4.5 a.u., 
  (c) low velocity of 4.5 a.u. with the azimuthal angle of 22$^\circ$.}
\label{fig:Lz_visual}
\end{figure*}

For a general non-symmetric case without GP, 
\bea \notag
    \tau_{12}  &=&-\frac{\gamma^{-1}\Delta x\Delta y-\gamma(\Delta x-\beta\Delta y)(\Delta y+\beta\Delta x)}
    {2[\gamma^{-1}\Delta x^2+\gamma(\Delta y+\beta \Delta x)^2]^2} \notag\\ 
    && - \frac{i}{2}[\gamma^{-1}\Delta x^2+\gamma(\Delta y+\beta \Delta x)^{2}]^{-1}L_z 
\label{eq:NAC_gen}
\eea
and this expression does not allow for a simple analysis of non-transferable terms. Therefore, we consider a symmetric model with $\beta=0$ and arbitrary $\gamma$. In this case, $\tau_{12}$ can be written as
\begin{equation}\label{eq:NAC_2D}
\begin{split}
       \tau_{12} & = \frac{(\gamma-\gamma^{-1})\cos\phi\sin\phi}{2r^2(\gamma^{-1}\cos\phi^2+\gamma \sin\phi^2)^2} \\
                 &- \frac{1}{2r^2(\gamma^{-1}\cos\phi^2+\gamma \sin\phi^2)}\frac{\partial}{\partial \phi}. \\
\end{split}
\end{equation}
By separation of variables, it is possible to solve a differential equation  
\begin{equation}
    \tau_{12}\chi_{\rm nt}(r,\phi) = 0
\end{equation}
for a non-transferable part of a nuclear wave-packet 
\begin{equation}
    \chi_{\rm nt}(r,\phi) = R(r)\sqrt{\gamma\sin^2\phi+\gamma^{-1}\cos^2\phi},
\end{equation}
where $R(r)$ is an arbitrary $r$-function independent of $\phi$. 
This component can be transferred when the GP is added because the corresponding 
$\tau_{12}^{\rm (GP)}$ acquires more terms
\begin{equation}\label{eq:NAC_GP_2Dg}
\begin{split}
\tau_{12}^{\text{(GP)}} = &  \frac{(\gamma-\gamma^{-1})\sin\phi\cos\phi}{2r^2(\gamma^{-1}\cos^2\phi+\gamma \sin^2\phi)^2}+\frac{1}{2ir^2(\gamma^{-1}x^2+\gamma y^2)}L_z\\ & +\frac{1}{8r^2(\gamma^{-1}\cos^2\phi+\gamma \sin^2\phi)^2}.
\end{split}
\end{equation}
However, it appears that the non-transferable component $\chi_{\rm nt}(r,\phi)$ for general $\gamma$ provides a less intuitive form than its simpler counterpart restricted to $\gamma=1$. Intuitive simplicity of the $\gamma=1$ analysis stems from its connection with classical mechanics. It is easy to estimate the weight of the $m=0$ component for a quantum wave-packet by considering it as an ensemble of classical trajectories and their classical angular momenta ($L_z=r\times p$) with respect to the CI. It was found that discretized classical estimates accurately represent the quantum weights for different $m$'s.\cite{Gherib:2015/jctc/11} This approach is illustrated in \fig{fig:Lz_visual}: for the head-on collisions the weight of $m=0$ component is reducing with the momentum of the wave-packet, for wave-packets traveling on a side of the CI the weight of $m=0$ component is reduced even further.

\section{Results and Discussion}
\label{sec:res_disc}

In this section, we evaluate the GP importance for the nonadiabatic population transfer between adiabatic states. 
The initial state is the ground vibrational state of the 1st diabatic state 
\bea
\chi(x,y,0) = \left(\frac{\omega_x\omega_y}{\pi^2}\right)^{1/4}e^{-\omega_x(x-x_0)^2/2-\omega_yy^2/2}.
\eea
placed vertically to the excited adiabatic state.

We investigate how breaking the subsystem Hamiltonian ($H_S$) symmetry by changing $\Delta_{12}$, $\Delta$, and $c_x$
affects the significance of GP effects. 
Dynamics of the excited electronic adiabatic state population will serve as the main dynamical indicator. Monitoring 
differences in population dynamics 
of exact [\eq{eq:H2D}], ``no GP'' [\eq{eq:Hadi}], and ``no GP no DBOC'' [\eq{eq:Hadi} with $\tau_{ii}=0$] 
systems are intended to illustrate the \gls{GP} effects.
The exact dynamics is simulated with the split-operator method in the diabatic basis with \eq{eq:H2D},
while both cases without \gls{GP} are done using the Chebyshev's polynomial expansion in the adiabatic basis~\cite{Tannor:2007}.  The CI point has been excluded from the space grid to avoid numerical issues 
related to divergences in the nonadiabatic coupling terms.
Weights of the $m=0$ and $m\ne 0$ components of a wave-packet on the excited state are evaluated as 
\bea
w_{m=0} &=& \int_0^{\infty} r|C_0(r)|^2 dr \\
w_{m\ne 0} &=& 1- w_{m=0}.
\eea
To illustrate the transition enhancement effect of $m=0$ component, we set $\omega_x=\omega_y=2$, $x_0=\frac{3}{2}$, $c_y=4$ and $\gamma = 0.33$.
To expose the \gls{DBOC} compensation effect, we consider a low kinetic energy configuration with $\omega_x=\omega_y=0.2$, $x_0=\frac{5}{2}$, $c_y=0.02$ and $\gamma = 0.2$.
Other parameters are set to zero if their values are not specified.

\subsection{Constant Coupling}

\begin{figure}[ht!]
  \centering
    \includegraphics[width=0.5\textwidth]{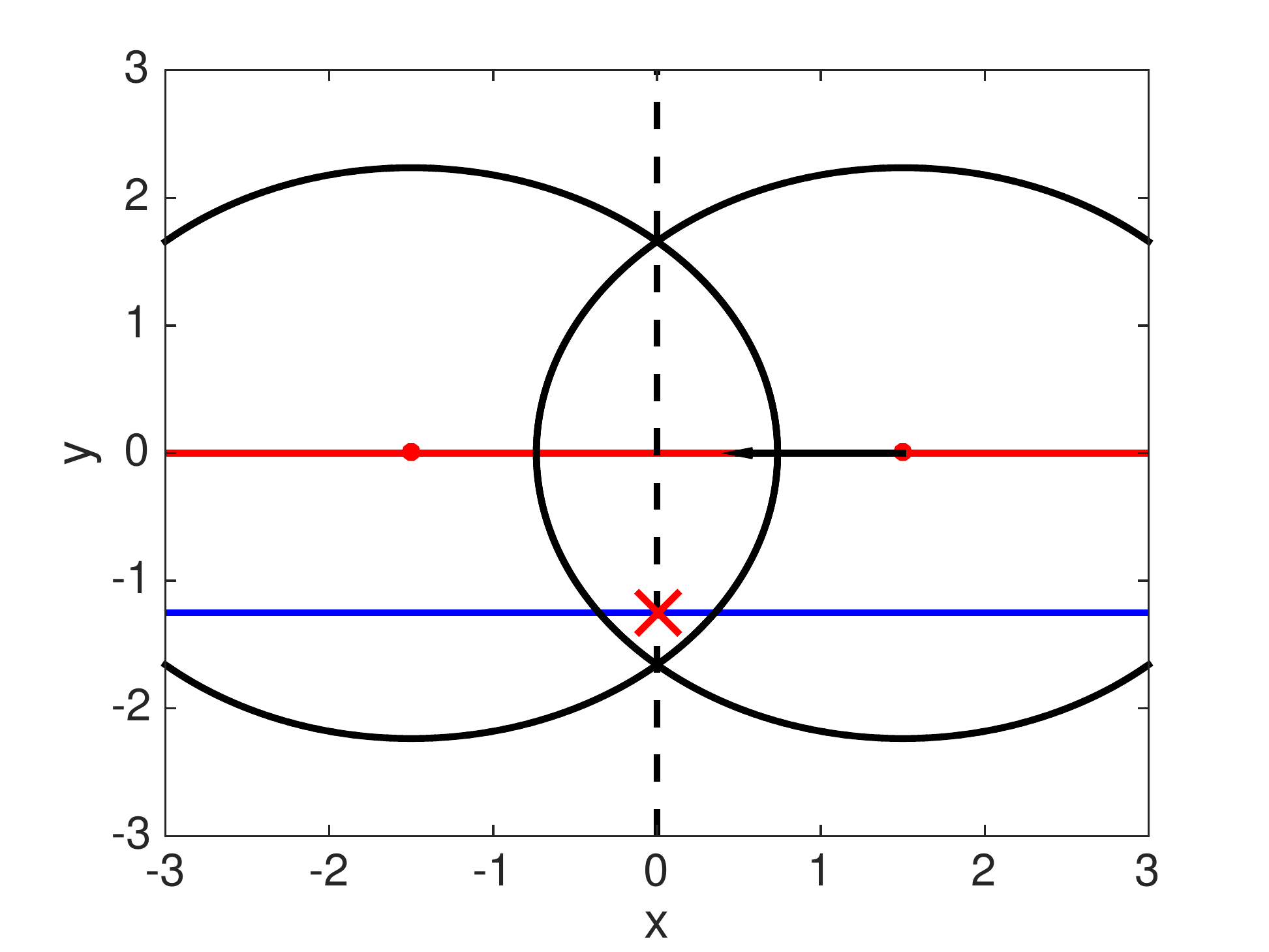}
  \caption{Top view of a system with $\Delta_{12}\neq 0$. 
  Meaning of the lines is the same as on \fig{fig:paramgeom}. 
  The arrow denotes the direction of the wave-packet motion.}
\label{fig:geom-cst-cpl}
\end{figure}

\begin{figure}[ht!]
  \centering
    \includegraphics[width=0.8\textwidth]{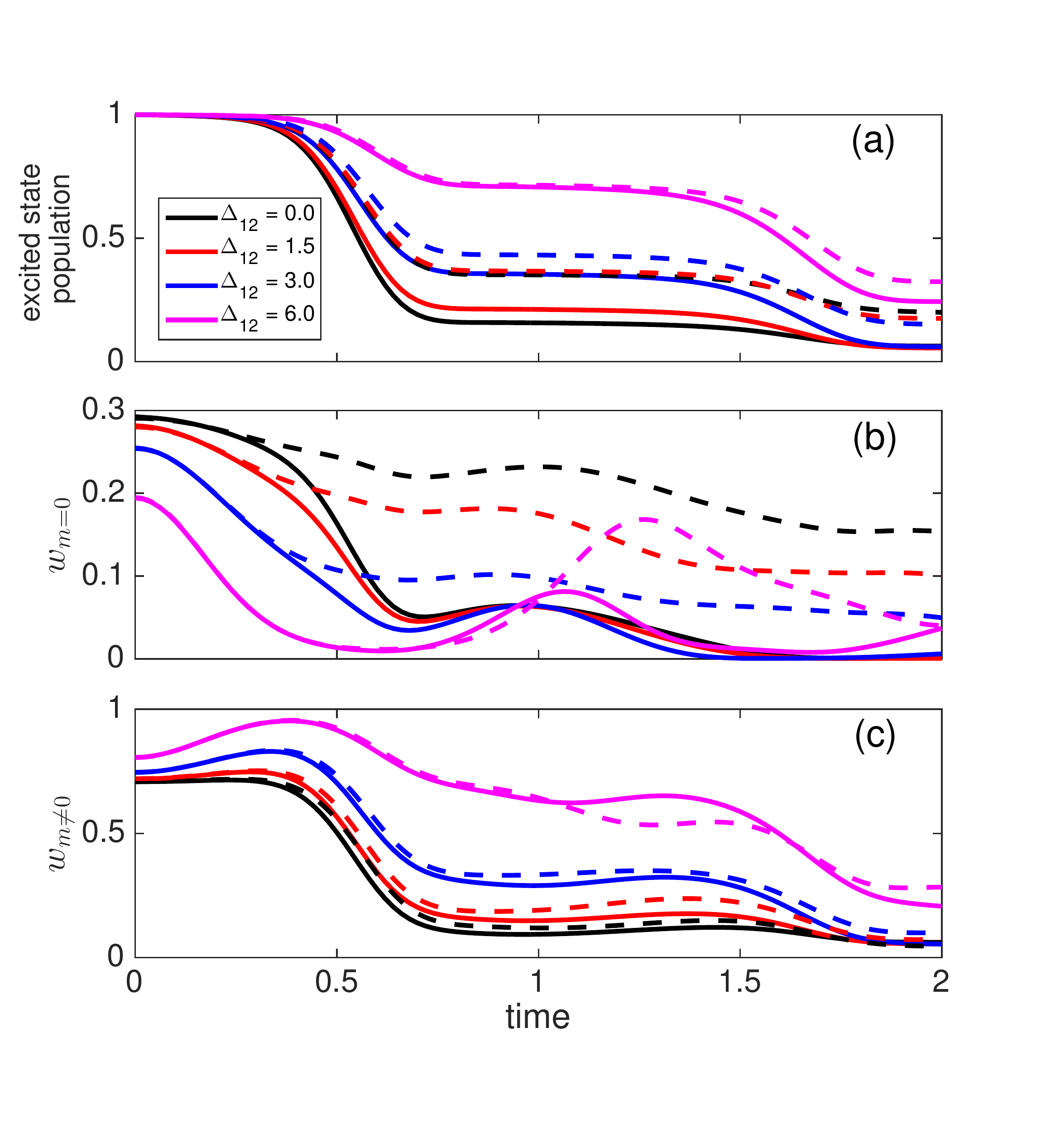}
  \caption{Dynamics with exact (solid) and ``no GP'' (dashed) Hamiltonians for systems with different $\Delta_{12}$'s: 
  (a) excited state population, (b) $m=0$ component weight, (c) $m\neq 0$ component weight.}
\label{fig:plot-cst-cpl}
\end{figure}

Non-zero $\Delta_{12}$ shifts the CI from the origin to $(0,-\Delta_{12}/c_y)$  (\eq{eq:ci_coor}). 
Therefore, when the wave-packet  moves along the tuning direction (\fig{fig:geom-cst-cpl}),
we can expect a decrease of $w_{m=0}$ (\fig{fig:Lz_visual} right). Thus, the transfer enhancement due to the 
GP will decrease with increase of $|\Delta_{12}|$. This is clearly confirmed in \fig{fig:plot-cst-cpl}a. 

By monitoring the value of $w_{m=0}$ (\fig{fig:plot-cst-cpl}b), 
one can see that ``no GP'' systems have larger $m=0$ component
weights than their exact counterparts, especially when nonadiabatic transfer is significant. 
On the other hand, $w_{m\ne 0}$ dynamics shows that the GP does not significantly alter 
nonadiabatic transfer rates for the $m\ne 0$ components (\fig{fig:plot-cst-cpl}c).

\begin{figure}[ht!]
  \centering
    \includegraphics[width=0.5\textwidth]{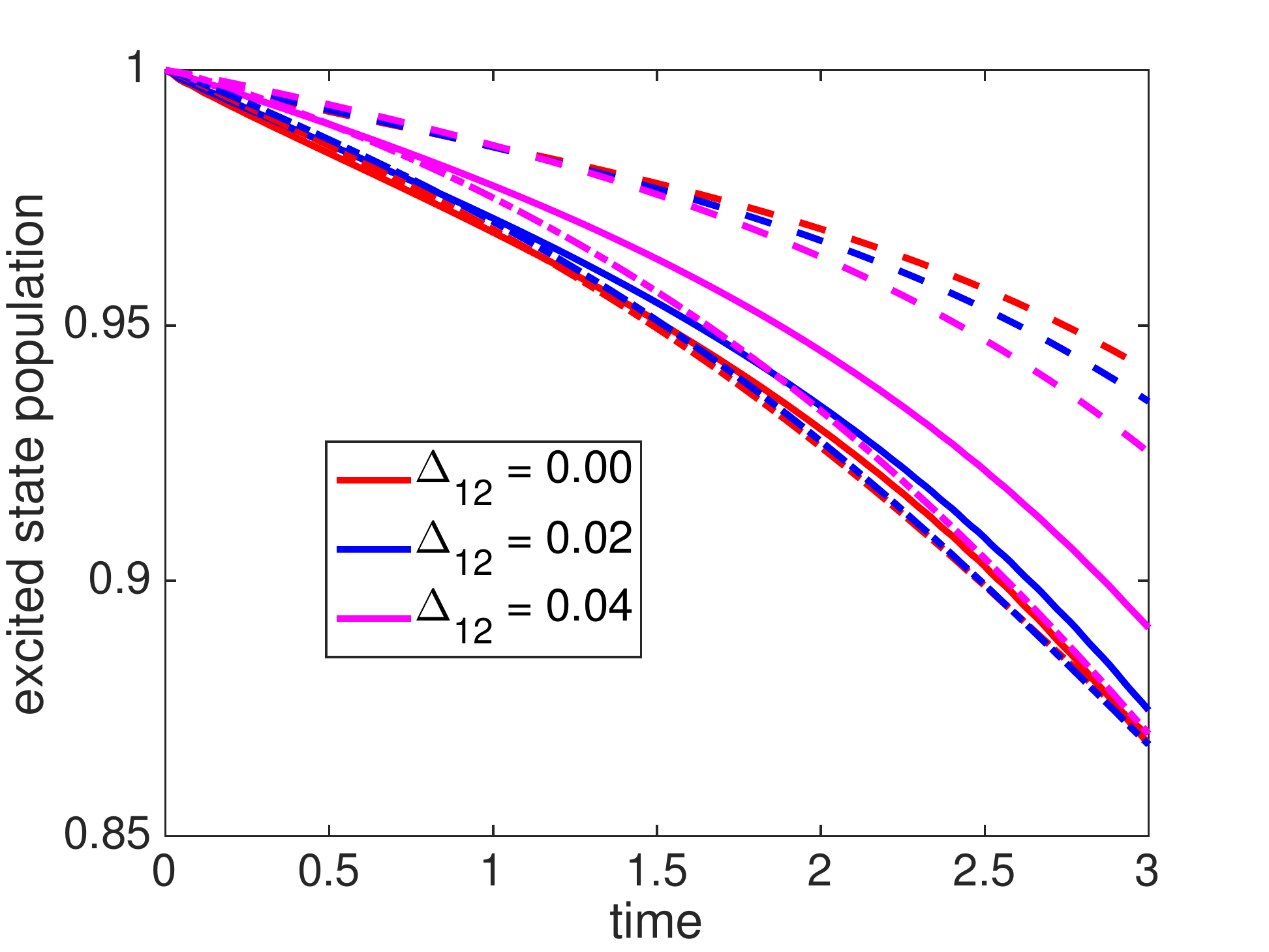}
  \caption{Population dynamics in the low kinetic energy model for various $\Delta_{12}$'s and with different Hamiltonians: 
  exact (solid), ``no \gls{GP}'' (dashed) and ``no \gls{GP} no \gls{DBOC}'' (dash-dot).}
\label{fig:c_0_02_d12}
\end{figure}

For the low energy configuration, dynamics and transfer are both significantly slower  (\fig{fig:c_0_02_d12}). 
Removing the \gls{DBOC} clearly helps to enhance transfer in the absence of \gls{GP}.
The \gls{DBOC} compensation by GP is most important when $\Delta_{12}$ is small (\fig{fig:c_0_02_d12}). 
Since the wave-packet moves close to the $y=0$ line, the non-zero constant coupling is the main contributor to the overall 
$V_{12}$ coupling. Hence, the wave-packet encounters smaller \gls{DBOC} values since the DBOC is inversely proportional 
to the adiabatic gap between states. 
Geometrically, the constant coupling also shifts the \gls{DBOC} centre because it shifts the \gls{CI}. 
Without constant coupling, a significant portion of \gls{DBOC} is on the path of the wave-packet  and slows it down. 
When the \gls{DBOC} is shifted by $\Delta_{12}\neq0$, only a wing of the \gls{DBOC} is affecting the wave-packet .

\subsection{Potential Energy Difference}

\begin{figure}[ht]
  \centering
    \includegraphics[width=0.5\textwidth]{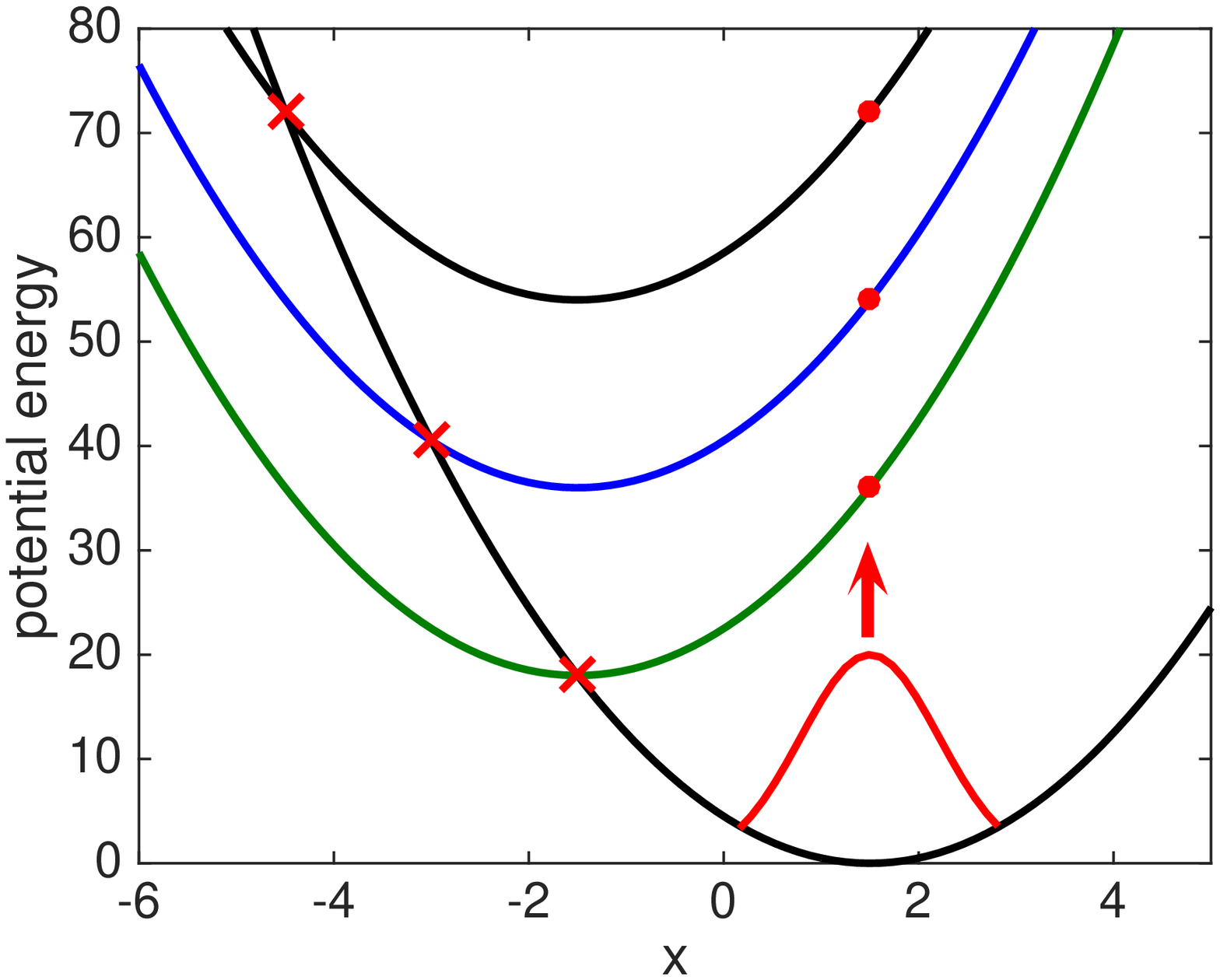}
  \caption{Cross sections of the diabatic surfaces at $y=0$,  the lowest minimum black curve for $V_{11}$
  and shifted to the left for $V_{22}$. $V_{22}$ cross sections are given for different $\Delta$'s: 
  $\Delta=18$ (green), $\Delta=36$ (blue), $\Delta=54$ (black). 
  The crosses and dots denote the CIs and FC points, respectively.}
\label{fig:geom-Ediff}
\end{figure}

\begin{figure}[ht!]
  \centering
    \includegraphics[width=0.8\textwidth]{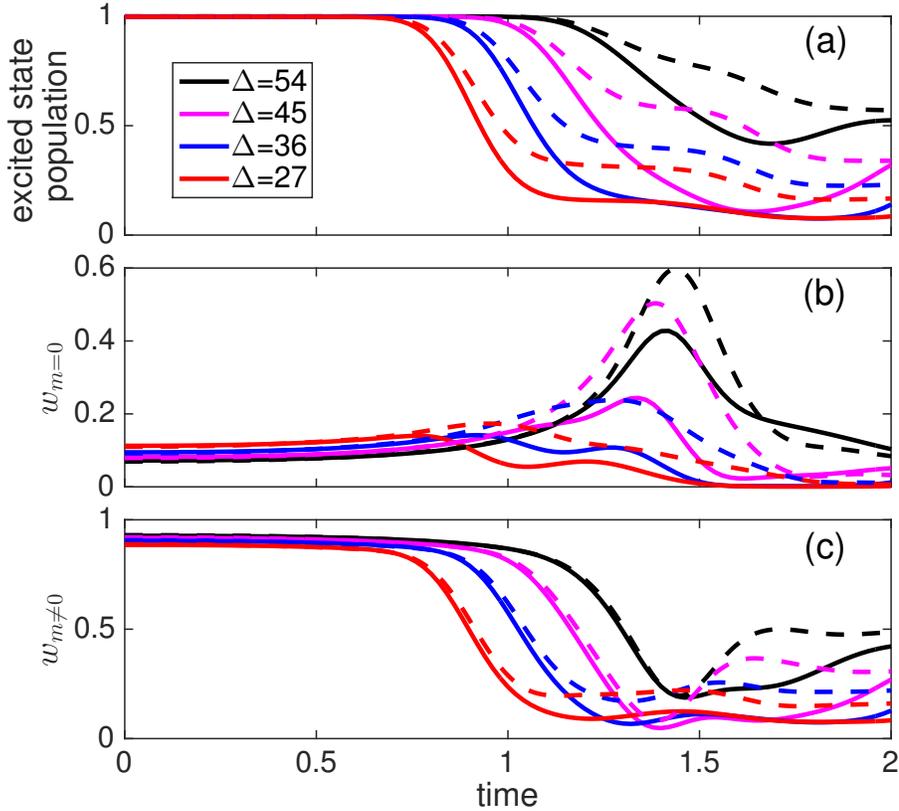} 
  \caption{Dynamics with exact (solid) and ``no GP'' (dashed) Hamiltonians for systems with energy differences $\Delta$: 
  (a) excited state population, (b) $m=0$ component weight, (c) $m\neq 0$ component weight.}
\label{fig:plot-Ediff}
\end{figure}

Variations of the energy difference, $\Delta$, move the degeneracy line along the tuning direction and 
can modify the character of the CI from peaked to sloped (\fig{fig:geom-Ediff}). 
This model has the reflective symmetry with respect to the $x$-axis. Hence, the wave-packet  is always 
moving along the tuning direction. Unlike in the $\Delta_{12}\ne 0$ case, here, the wave-packet collides 
with the CI head-on for all values of $\Delta$. 
$\Delta=18$ corresponds to the highest kinetic energy of the 
wave-packet at the CI, while $\Delta=54$ makes the Franck-Condon (FC) and CI points equal in energy.
Wave-packets with smaller linear momenta at the collision moment
have higher $m=0$ component weights. Therefore, reducing the energy difference between the FC   
and CI points can emphasize the GP effect on population transfer (\fig{fig:plot-Ediff}a). 
One seeming exception from this trend is $\Delta=54$, where the wave-packet has no kinetic energy when it arrives at the CI. 
In this case, the wave-packet does not pass through the CI twice like the wave-packets for other values of $\Delta$, 
but moves back and accelerates toward the FC point. This results in reduced transfer and somewhat less pronounced GP effects
as compared to $\Delta=45$.

Expectedly, the $m=0$ weights grow when wave-packets approach the CI from the FC point (\fig{fig:plot-Ediff}b).
Due to higher transfer rates in the exact dynamics, 
$m=0$ weights of excited state wave-packets peak at higher values in the ``no GP'' case rather than in the exact dynamics.
For $m\neq0$ components, the difference between simulations with and without GP is negligible up to the time 
when the wave-packets are moving away from the CI (\fig{fig:plot-Ediff}c). 
The motion from the CI converts the residual $m=0$ component to the $m\neq0$ part 
due to the increasing linear momentum.

\begin{figure}[ht!]
  \centering
    \includegraphics[width=0.5\textwidth]{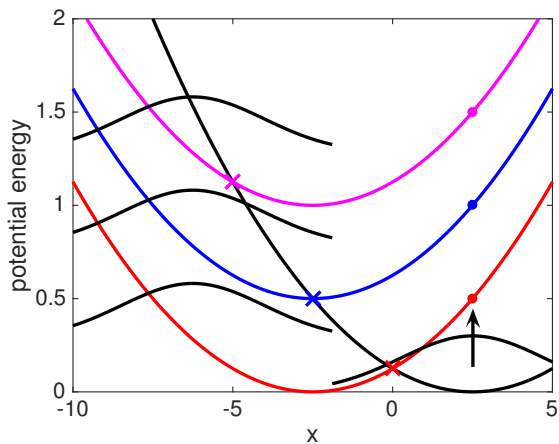}
  \caption{Cross sections of the diabatic surfaces at $y=0$ for the low kinetic energy model, $V_{11}$ (black) 
  and $V_{22}$ (coloured): red is for $\Delta=0$ with a peaked CI,
  blue is for $\Delta=0.5$ with the CI at the minimum of $V_{22}$, and magenta is for $\Delta=1.0$ with the CI not located between the diabatic minima.
  The crosses and dots denote the CIs and FC points, respectively. 
  The excited state Gaussians show schematically the nuclear 
  wavefunction at the turning points.
}
\label{fig:D2}
\end{figure}

\begin{figure}[ht!]
  \centering
    \includegraphics[width=0.5\textwidth]{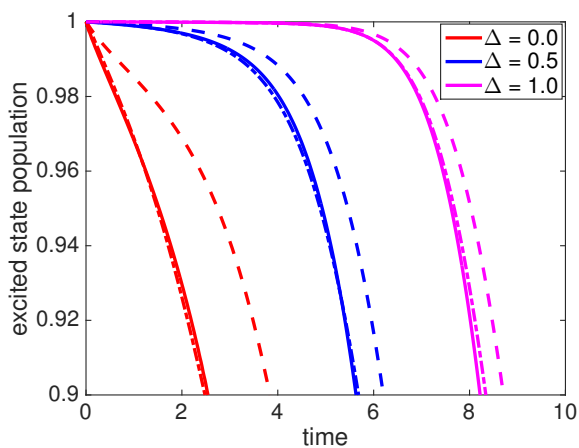}
  \caption{Population dynamics with the low kinetic energy model and different values of $\Delta$:  
  exact (solid), ``no \gls{GP}'' (dashed), and ``no \gls{GP} no \gls{DBOC}'' 
  (dash-dot).}
\label{fig:c_0_2_d}
\end{figure}

To study compensation of the \gls{DBOC} by \gls{GP} we considered the low kinetic energy model 
and rescaled values of $\Delta$ as illustrated in \fig{fig:D2}. 
The \gls{DBOC} compensation is observed in population dynamics for all cases (\fig{fig:c_0_2_d}), 
and it is the most prominent for $\Delta = 0$. 
In this case, the \gls{CI} is peaked and the entire wave-packet goes through the \gls{DBOC} barrier in exact dynamics (\fig{fig:D2}). 
However, without GP, it does not have enough kinetic energy to overcome the \gls{DBOC} barrier, 
which decreases the population transfer. 
Increasing $\Delta$ leads to reducing the DBOC significance, this can be related to the portion of a nuclear wave-packet that access the CI and is strongly affected by the DBOC (\fig{fig:D2}).   
The reduction of this portion stems from the repulsive nature of the accepting $W_{-}$ potential that forces the wave-packet to 
turn back. Interestingly, this trend can be reformulated as a general reduction of the DBOC influence for the sloped CIs compared 
to the peaked counterparts. 

\subsection{Nonorthogonal linear coupling}

\begin{figure}[ht]
  \centering
    \includegraphics[width=0.5\textwidth]{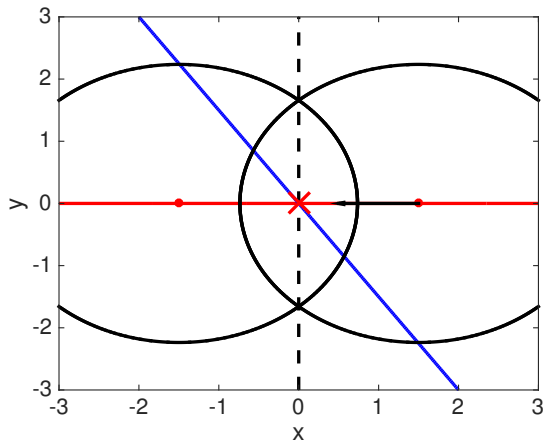}
  \caption{Top view of a non-orthogonal system.
   Objects in the figure have the same meaning as in \fig{fig:geom-cst-cpl}.}
\label{fig:geom-nonortho}
\end{figure}

The linear coupling $V_{12}$ is considered to be orthogonal 
if the zero-coupling line is orthogonal to the degeneracy line, which is equivalent to $\beta=0$. 
When $\beta\ne 0$, the linear coupling $V_{12}$ is nonorthogonal (\fig{fig:geom-nonortho}), 
as a measure of nonorthogonality we will use a deviation of the angle between the zero-coupling and 
degeneracy lines from 90 degrees. This deviation is given by $\varphi=\arctan(\beta)$.
For nonorthogonal cases, $\tau_{12}$ cannot be expressed as in \eq{eq:NAC_GP_2D}. 
Instead, considering \eq{eq:NAC_gen} 
suggests that even without GP, an extra channel of nonadiabatic transfer opens for the $m=0$ component. 
Nevertheless, since the $L_z$-dependent part still exists, the GP can enhance transfer of the $m=0$ component as well.

\begin{figure}[ht!]
  \centering
    \includegraphics[width=0.8\textwidth]{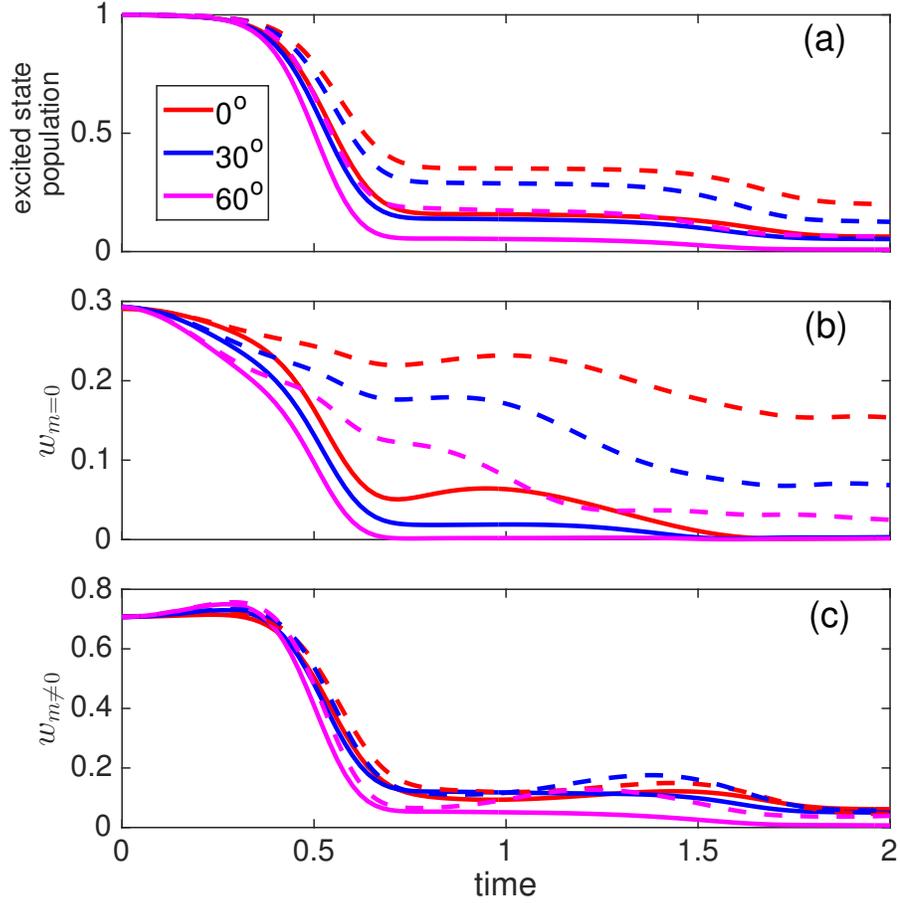}
  \caption{Dynamics with exact (solid) and ``no GP'' (dashed) Hamiltonians for 
  orthogonal and non-orthogonal systems: (a) excited state population, (b) $w_{m=0}$,  (c) $w_{m\neq0}$. }
\label{fig:nonorth_pop}
\end{figure}

Population results for different angles ($\varphi$) show that increasing the degree of nonorthogonality reduces
the importance of the GP (\fig{fig:nonorth_pop}a). 
This observation can be rationalized by noticing that
nonorthogonality of the diabatic coupling breaks the reflective symmetry with respect to the $x$-axis 
for the adiabatic PESs, $W_{\pm}$. 
A wave-packet starting at the FC point on $W_{+}$ will not move toward the CI in a straight line. 
This change of the trajectory affects values of $w_{m=0}$ and $w_{m\ne 0}$ similarly 
to the $\Delta_{12}\neq 0$ case. 
Although $m=0$ component transfer is not forbidden if $\gamma\neq 1$ or $\beta\neq 0$, \fig{fig:nonorth_pop}b suggests that the transition enhancement from the GP is still crucial for moderate $\varphi$. As in other cases, transfer rates of $m\neq0$ components are unaffected by GP contributions (\fig{fig:nonorth_pop}c). 

\begin{figure}[ht!]
  \centering
  \includegraphics[width=0.5\textwidth]{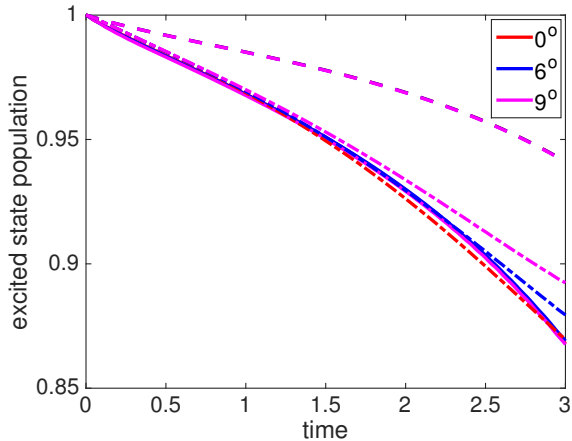}
\caption{Population dynamics of orthogonal and non-orthogonal systems in the 
  exact(solid)  ``no GP''(dash)  and ``no GP no DBOC''(dash-dot) approaches. All ``no GP'' and exact dynamics results 
  are visually indistinguishable for different angles.}
\label{fig:testDBOC}
\end{figure}

For $\beta\neq 0$ and low $\gamma$, the \gls{DBOC} ridge forms an angle with the $x$-axis (see \fig{fig:DBOCshape}). 
Unfortunately, values of $\varphi$ that do not lead to numerical instabilities in the low kinetic energy model, 
are too small to produce visible changes within ``no GP'' or exact dynamics. Thus, from 
\fig{fig:testDBOC} it is impossible to predict how $\varphi$ increase will affect the GP influence. 
In the accessible range of $\varphi$, the \gls{DBOC} removal increases transfer to the 
level of the exact dynamics (\fig{fig:testDBOC}). 

\section{Conclusions}
\label{sec:conc}

In this work we systematically investigated influence of symmetry breaking on GP effects 
in excited state nonadiabatic dynamics of the N-dimensional LVC model. 
The proposed analysis can be used to estimate how important GP effects are in 
systems for which ND-LVC can be obtained.  The core of our analysis is in modelling nonadiabatic 
dynamics of the effective 2D subsystem that is affected by all parameters of the original ND-LVC model. 
In all examples breaking symmetry in the effective 2D subsystem reduces the importance of GP.
This is easy to understand considering that GP effects are pronounced if at least one of the
 two conditions is satisfied, either the DBOC is prominent because of low coupling strength $\gamma$ 
 or the non-rotating ($m=0$) component of the nuclear wave-packet is significant in the CI vicinity. 
 Adding more nuclear DOF to the system generally decreases chances to satisfy one or both of the conditions.
This confirms that although GP is a topological phenomenon, its role in molecular dynamics
strongly depends on the energy landscape and the path of the nuclear wave-packet with respect to 
the CI seam.

\section{Acknowledgements}
J.L. is grateful to University of Toronto for financial support through an Excellence Award. A.F.I. acknowledges funding from a Sloan Research Fellowship and the Natural Sciences and Engineering Research Council of Canada (NSERC) through the Discovery Grants Program.

\appendix

\section{Reduced dimensional Hamiltonian model}
\label{app:transfo}

Short-time dynamics of the \gls{LVC} model can be captured within reduced dimensional models.~\cite{Cederbaum:2005/prl/113003,Gindensperger:2006/jcp/144103,Gindensperger:2006/jcp/144104,Ryabinkin:2014/jcp/214116} In our previous work, we devised such a reduction to two-dimensional two-states Hamiltonians that are capable of reproducing short-time diabatic population dynamics of the full-dimensional model.~\cite{Ryabinkin:2014/jcp/214116} This is possible because the reduced Hamiltonian is defined so that its first moments of the cumulant expansion for the population dynamics are the same as those of the full-dimensional model. The reduced Hamiltonian is obtained by defining two sets of collective coordinates: two subsystem coordinates and remaining bath coordinates. This definitions allow us to rewrite the total N-dimensional model Hamiltonian a subsystem-bath Hamiltonian. Here, we demonstrate how to obtain parameters of the subsystem-bath model from those of the N-dimensional model [\eq{eq:HND}].

\subsection{Transformation to the reduced model}

Obtaining the subsystem-bath Hamiltonian in \eq{eq:H2D} from that of \eq{eq:HND} can be done in three steps.

{\it Step 1:} Equation \eqref{eq:HND} is rewritten in frequency-weighted nuclear coordinates
\bea
\tilde q_j=\sqrt{\Omega_j}q_j.
\eea
For our choice of initial conditions, $\tilde\kappa_j=0$, the transformed Hamiltonian is
\bea \notag
    \boldsymbol{H}_1 &=& \sum_{j}^{N}\frac{\Omega_j}{2}(\tilde p_j^2 + \tilde q_j^2)\boldsymbol{I}_2 + \begin{bmatrix} 0 & L_j\tilde q_j \\ L_j\tilde q_j & K_j\tilde q_j \end{bmatrix}+ \begin{bmatrix} -{\delta/2} & 0 \\ 0 & {\delta/2} \end{bmatrix}, \\ \label{eq:HND2}
\eea
where the new parameters are defined as 
$K_j = \kappa_j/\sqrt{\Omega_j}$, $L_j = \lambda_j/\sqrt{\Omega_j}$, 
and $\tilde p_j = p_j/\sqrt{\Omega_j}$.

{\it Step 2:} The nuclear coordinates $\tilde{q}_j$ are rotated to new collective coordinates
\bea
(\tilde X, \tilde Y, \tilde Q_1, \tilde Q_2\dots)^{T}=\boldsymbol T\tilde{\boldsymbol q}.
\eea
The linear transformation $\boldsymbol T$ is defined as a product of two rotations: $\boldsymbol T = \boldsymbol U\boldsymbol R$. The first transformation, $\boldsymbol R$, is used to define subsystem and bath collective coordinates and is expressed as $\boldsymbol R = [ \boldsymbol R_S^T \boldsymbol R_B^T]^T$, where 
\bea
\boldsymbol R_S & = & \begin{bmatrix}\sum_j K_j^2&\sum_j K_jL_j\\\sum_j K_jL_j&\sum_j L_j^2\end{bmatrix}^{-1/2}\begin{bmatrix}\boldsymbol K^T\\\boldsymbol L^T\end{bmatrix}
\eea
rotates and projects $\boldsymbol{\tilde q}$ in the subsystem subspace while $\boldsymbol R_B$ projects onto bath subspace. Rows of $\boldsymbol R_B$ are obtained by Gram-Schmidt orthonormalization of the ND-space basis with respect to the 2 rows of $\boldsymbol R_S$. The second transformation $\boldsymbol U$ diagonalizes the frequency matrices in the subsystem, $\boldsymbol R_S^T\boldsymbol\Omega\boldsymbol R_S$, and bath, $\boldsymbol R_B^T\boldsymbol\Omega\boldsymbol R_B$, subspaces
\bea
 \boldsymbol U = (\begin{smallmatrix}\boldsymbol U_S&\boldsymbol 0\\\boldsymbol 0&\boldsymbol U_B\end{smallmatrix})
\eea 
and satisfies
\bea
\boldsymbol U_S\boldsymbol R_S\boldsymbol \Omega\boldsymbol R_S^T\boldsymbol U_S^T & = & \mathrm{diag}(\omega_x\,\omega_y), \\
\boldsymbol U_B\boldsymbol R_B\boldsymbol \Omega\boldsymbol R_B^T\boldsymbol U_B^T & = & \mathrm{diag}(\tilde\Omega_1\,\tilde\Omega_2\,\dots),
\eea
where $\omega_x$, and $\omega_y$ are the subsystem frequencies, and $\{\tilde\Omega_j\}$ are the bath frequencies. Applying transformation $\boldsymbol T$ to $\boldsymbol{H}_1$ gives a Hamiltonian which consists of three terms:
\bea\label{eq:H2}
\boldsymbol{H}_2 & = & \boldsymbol{H}_{2,S} + \boldsymbol{H}_{2,B} + \boldsymbol{H}_{2,SB}. 
\eea
The subsystem Hamiltonian is
\bea\label{eq:H2S}
\boldsymbol{H}_{2,S} & = & \frac{\omega_x}{2}\left(\tilde X^2-\partial^2/\partial \tilde X^2\right)\boldsymbol{I}_2 + \frac{\omega_y}{2}\left(\tilde Y^2-\partial^2/\partial \tilde Y^2\right)\boldsymbol{I}_2 \nonumber\\
&&+ \begin{bmatrix} 0 & \tilde c_x\tilde X+\tilde c_y\tilde Y \\ \tilde c_x\tilde X+\tilde c_y\tilde Y & \tilde d_x\tilde X + \tilde d_y\tilde Y \end{bmatrix} + \begin{bmatrix} -{\delta/2} & 0 \\ 0 & {\delta/2} \end{bmatrix}.
\eea
where $\tilde{\boldsymbol d}=\boldsymbol U_S\boldsymbol R_S\boldsymbol K$ and $\tilde{\boldsymbol c}=\boldsymbol U_S\boldsymbol R_S\boldsymbol L$ ($||\boldsymbol R_B\boldsymbol K||=||\boldsymbol R_B\boldsymbol L||=0$ by construction).
The bath Hamiltonian is
\bea\label{eq:H2B}
\boldsymbol{H}_{2,B} & = & \sum_{j=1}^{N-2}\frac{\tilde\Omega_j}{2}\left(\tilde P_j^2+\tilde Q_j^2\right)\boldsymbol{I}_2,
\eea
where $\tilde{\boldsymbol P}=\boldsymbol U_B\boldsymbol R_B\tilde{\boldsymbol p}$.
The subsystem-bath coupling is
\bea\label{eq:H2SB}
\boldsymbol{H}_{2,SB} & = & \sum_{j=1}^{N-2} \bigg[\Lambda_{jx}\bigg(\tilde X\tilde Q_j +  \frac{\tilde P_j}{i}\frac{\partial}{\partial\tilde X} \bigg) \nonumber\\
&&\hspace{1cm}+\Lambda_{jy}\bigg( \tilde Y\tilde Q_j + \frac{\tilde P_j}{i}\frac{\partial}{\partial\tilde Y} \bigg)\bigg] \boldsymbol{I}_2, 
\eea
where the couplings are defined as
\bea\label{eq:Lambdas}
\Lambda_{jx} & = & \sum_{k=1}^N \Omega_kT_{j+2\,k}T_{1k} \\
\Lambda_{jy} & = & \sum_{k=1}^N \Omega_kT_{j+2\,k}T_{2k}.
\eea

{\it Step 3:} The final subsystem-bath Hamiltonian in \eq{eq:Htot} is obtained from \eq{eq:H2} by applying an affine transformation that removes the frequency weighting and shifts the coordinates
\bea
(x\, y\, Q_1\, Q_2\,\dots)^{T}=\boldsymbol \omega^{-1/2}(\tilde X, \tilde Y, \tilde Q_1, \tilde Q_2\dots)^{T} + \boldsymbol \omega^{-3/2}\boldsymbol d/2.
\eea
The system Hamiltonian in the new coordinates is given by \eq{eq:Htot} in Sec.~\ref{sec:theory}. All quantities for this Hamiltonian are obtained as follows
\bea
\boldsymbol\nabla & = & \begin{pmatrix}\sqrt{\omega_x}\partial/\partial\tilde X,&\sqrt{\omega_y}\partial/\partial\tilde Y\end{pmatrix}^T, \\
P_j & = & \tilde P_j/\sqrt{\tilde\Omega_j}, \\
\begin{pmatrix}x_0,&y_0\end{pmatrix} & = & \begin{pmatrix}\omega_x^{-3/2}\tilde d_x/2, &\omega_y^{-3/2}\tilde d_y/2\end{pmatrix}, \\
\begin{pmatrix}c_x, &c_y\end{pmatrix} & = & \begin{pmatrix}\sqrt{\omega_x}\tilde c_x, &\sqrt{\omega_y}\tilde c_y\end{pmatrix}, \\
f_j & = & -\sqrt{\tilde\Omega_j}\bigg[\frac{\Lambda_{jx}\tilde d_x}{\sqrt{\omega_x}}+\frac{\Lambda_{jy}\tilde d_y}{\sqrt{\omega_y}}\bigg], \\
\Delta & = & \delta-\frac{1}{2}\big[\tilde d_x^2/\omega_x+\tilde d_y^2/\omega_y\big], \\
\Delta_{12} & = & -\frac{1}{2}\big[\tilde c_x\tilde d_x/\omega_x+\tilde c_y\tilde d_y/\omega_y\big].
\eea
The final coordinate transformation gives rise to an energy shift term $-\sum_k\tilde d_k^2/2\omega_k\boldsymbol I_2$, which acts identically for both states and thus does not affect dynamics.

\subsection{Subsystem-bath effect}

The subsystem Hamiltonian in \eq{eq:H2S} can reproduce the short time population dynamics of the full Hamiltonian \eq{eq:H2}.~\cite{Ryabinkin:2014/jcp/214116} Expanding the population of the diabatic state two up to the fourth order in Taylor series with respect to time gives for \eq{eq:H2}
\bea\label{eq:P2tS}
P_{2,S}(t) & = & 1 - \frac{t^2}{2}\sum_j \tilde c_j^2 + \frac{t^4}{24}\bigg(\sum_{j}^{x,y}\tilde c_j\tilde d_j\bigg)^2 \nonumber\\
&&+ \frac{t^4}{12}\sum_{j}^{x,y}\tilde c_j^2\bigg[(\omega_j-\delta)^2+\sum_l^{x,y}\Big(3\tilde c_l^2+\frac{\tilde d_l^2}{4}\Big)\bigg] \nonumber\\
&&+\frac{t^4}{12}\sum_{j,k}^{x,y}\tilde c_j\tilde c_k\sum_l^{N-2}\Lambda_{lj}\Lambda_{lk} + \mathcal{O}(t^6)
\eea
The subsystem-bath couplings $\Lambda_{lk}$ appear in the population dynamics only in the the fourth order, while the first non-trivial term depends only on the subsystem parameters. Although, higher order contributions are too cumbersome to evaluate explicitly, it can be shown that they have terms proportional to $\sum_{j,k,l}\tilde c_j\tilde d_k\Lambda_{lj}\Lambda_{lk}$ and $\sum_{j,k,l}\tilde d_j\tilde d_k\Lambda_{lj}\Lambda_{lk}$. These terms can be rewritten in a matrix form with parameters from \eq{eq:HND}:
\bea\label{eq:cplterms}
\sum_{j,k,l}\Lambda_{l,j}\Lambda_{l,k}\begin{bmatrix}\tilde d_j\tilde d_k &\tilde d_j\tilde c_k \\\tilde d_j\tilde c_k &\tilde c_j\tilde c_k \end{bmatrix} & = & \mathbf{M}_1 - \mathbf{M}_0\mathbf{M}_{-1}^{-1}\mathbf{M}_0, \nonumber\\
\eea
where matrices $\mathbf{M}_n$ are given by
\bea
 \mathbf{M}_n & = & \sum_{i=1}^N \Omega_i^{n} \begin{bmatrix} \kappa_i^2 & \kappa_i\lambda_i\\\kappa_i\lambda_i & \lambda_i^2 \end{bmatrix}.
\eea
The strength of the couplings can be estimated by the ratio
\bea
\frac{2||\boldsymbol\Lambda||^2}{||\boldsymbol\omega||^2+||\tilde{\boldsymbol\Omega}||^2} = \frac{2\sum_{i=1}^{N-2} (\Lambda_{ix}^2+\Lambda_{iy}^2)}{\omega_x^2+\omega_y^2+\sum_{j=1}^{N-2}\tilde\Omega_j^2}.
\eea
To make this ratio much smaller than unity, using \eq{eq:cplterms}, we can arrive at 
\bea\label{eq:timescale}
\mathrm{Tr}\big\{ (\mathbf{M}_1 - \mathbf{M}_0\mathbf{M}_{-1}^{-1}\mathbf{M}_0) \mathbf{M}_{-1}^{-1} \big\} & \ll & \sum_{j=1}^{N}\Omega_j^2. 
\eea
In other words, the interaction with the bath can be neglected for timescales at which the subsystem evolves nonadiabatically if \eq{eq:timescale} is satisfied.


\begin{thebibliography}{32}%
\makeatletter
\providecommand \@ifxundefined [1]{%
 \@ifx{#1\undefined}
}%
\providecommand \@ifnum [1]{%
 \ifnum #1\expandafter \@firstoftwo
 \else \expandafter \@secondoftwo
 \fi
}%
\providecommand \@ifx [1]{%
 \ifx #1\expandafter \@firstoftwo
 \else \expandafter \@secondoftwo
 \fi
}%
\providecommand \natexlab [1]{#1}%
\providecommand \enquote  [1]{``#1''}%
\providecommand \bibnamefont  [1]{#1}%
\providecommand \bibfnamefont [1]{#1}%
\providecommand \citenamefont [1]{#1}%
\providecommand \href@noop [0]{\@secondoftwo}%
\providecommand \href [0]{\begingroup \@sanitize@url \@href}%
\providecommand \@href[1]{\@@startlink{#1}\@@href}%
\providecommand \@@href[1]{\endgroup#1\@@endlink}%
\providecommand \@sanitize@url [0]{\catcode `\\12\catcode `\$12\catcode
  `\&12\catcode `\#12\catcode `\^12\catcode `\_12\catcode `\%12\relax}%
\providecommand \@@startlink[1]{}%
\providecommand \@@endlink[0]{}%
\providecommand \url  [0]{\begingroup\@sanitize@url \@url }%
\providecommand \@url [1]{\endgroup\@href {#1}{\urlprefix }}%
\providecommand \urlprefix  [0]{URL }%
\providecommand \Eprint [0]{\href }%
\providecommand \doibase [0]{http://dx.doi.org/}%
\providecommand \selectlanguage [0]{\@gobble}%
\providecommand \bibinfo  [0]{\@secondoftwo}%
\providecommand \bibfield  [0]{\@secondoftwo}%
\providecommand \translation [1]{[#1]}%
\providecommand \BibitemOpen [0]{}%
\providecommand \bibitemStop [0]{}%
\providecommand \bibitemNoStop [0]{.\EOS\space}%
\providecommand \EOS [0]{\spacefactor3000\relax}%
\providecommand \BibitemShut  [1]{\csname bibitem#1\endcsname}%
\let\auto@bib@innerbib\@empty
\bibitem [{\citenamefont {Truhlar}\ and\ \citenamefont
  {Mead}(2003)}]{Truhlar:2003/pra/032501}%
  \BibitemOpen
  \bibfield  {author} {\bibinfo {author} {\bibfnamefont {D.~G.}\ \bibnamefont
  {Truhlar}}\ and\ \bibinfo {author} {\bibfnamefont {C.~A.}\ \bibnamefont
  {Mead}},\ }\href {\doibase 10.1103/PhysRevA.68.032501} {\bibfield  {journal}
  {\bibinfo  {journal} {Phys. Rev. A}\ }\textbf {\bibinfo {volume} {68}},\
  \bibinfo {pages} {032501} (\bibinfo {year} {2003})}\BibitemShut {NoStop}%
\bibitem [{\citenamefont {Migani}\ and\ \citenamefont
  {Olivucci}(2004)}]{Migani:2004/271}%
  \BibitemOpen
  \bibfield  {author} {\bibinfo {author} {\bibfnamefont {A.}~\bibnamefont
  {Migani}}\ and\ \bibinfo {author} {\bibfnamefont {M.}~\bibnamefont
  {Olivucci}},\ }in\ \href@noop {} {\emph {\bibinfo {booktitle} {{Conical
  Intersection Electronic Structure, Dynamics and Spectroscopy}}}},\ \bibinfo
  {editor} {edited by\ \bibinfo {editor} {\bibfnamefont {W.}~\bibnamefont
  {Domcke}}, \bibinfo {editor} {\bibfnamefont {D.~R.}\ \bibnamefont {Yarkony}},
  \ and\ \bibinfo {editor} {\bibfnamefont {H.}~\bibnamefont {K\"{o}ppel}}}\
  (\bibinfo  {publisher} {World Scientific},\ \bibinfo {address} {New Jersey},\
  \bibinfo {year} {2004})\ p.\ \bibinfo {pages} {271}\BibitemShut {NoStop}%
\bibitem [{\citenamefont {Domcke}\ and\ \citenamefont
  {Yarkony}(2012)}]{Domcke:2012/arpc/325}%
  \BibitemOpen
  \bibfield  {author} {\bibinfo {author} {\bibfnamefont {W.}~\bibnamefont
  {Domcke}}\ and\ \bibinfo {author} {\bibfnamefont {D.~R.}\ \bibnamefont
  {Yarkony}},\ }\href {\doibase 10.1146/annurev-physchem-032210-103522}
  {\bibfield  {journal} {\bibinfo  {journal} {Annu. Rev. Phys. Chem.}\ }\textbf
  {\bibinfo {volume} {63}},\ \bibinfo {pages} {325} (\bibinfo {year}
  {2012})}\BibitemShut {NoStop}%
\bibitem [{\citenamefont {Yarkony}(1996)}]{Yarkony:1996/rmp/985}%
  \BibitemOpen
  \bibfield  {author} {\bibinfo {author} {\bibfnamefont {D.~R.}\ \bibnamefont
  {Yarkony}},\ }\href {\doibase 10.1103/RevModPhys.68.985} {\bibfield
  {journal} {\bibinfo  {journal} {Rev. Mod. Phys.}\ }\textbf {\bibinfo {volume}
  {68}},\ \bibinfo {pages} {985} (\bibinfo {year} {1996})}\BibitemShut
  {NoStop}%
\bibitem [{\citenamefont {Longuet-Higgins}\ \emph {et~al.}(1958)\citenamefont
  {Longuet-Higgins}, \citenamefont {Opik}, \citenamefont {Pryce},\ and\
  \citenamefont {Sack}}]{LonguetHigg:1958/rspa/1}%
  \BibitemOpen
  \bibfield  {author} {\bibinfo {author} {\bibfnamefont {H.~C.}\ \bibnamefont
  {Longuet-Higgins}}, \bibinfo {author} {\bibfnamefont {U.}~\bibnamefont
  {Opik}}, \bibinfo {author} {\bibfnamefont {M.~H.~L.}\ \bibnamefont {Pryce}},
  \ and\ \bibinfo {author} {\bibfnamefont {R.~A.}\ \bibnamefont {Sack}},\
  }\href {\doibase 10.1098/rspa.1958.0022} {\bibfield  {journal} {\bibinfo
  {journal} {Proc. R. Soc. A}\ }\textbf {\bibinfo {volume} {244}},\ \bibinfo
  {pages} {1} (\bibinfo {year} {1958})}\BibitemShut {NoStop}%
\bibitem [{\citenamefont {Berry}(1984)}]{Berry:1984/rspa/45}%
  \BibitemOpen
  \bibfield  {author} {\bibinfo {author} {\bibfnamefont {M.~V.}\ \bibnamefont
  {Berry}},\ }\href {\doibase 10.1098/rspa.1984.0023} {\bibfield  {journal}
  {\bibinfo  {journal} {Proc. R. Soc. A}\ }\textbf {\bibinfo {volume} {392}},\
  \bibinfo {pages} {45} (\bibinfo {year} {1984})}\BibitemShut {NoStop}%
\bibitem [{\citenamefont {Mead}\ and\ \citenamefont
  {Truhlar}(1979)}]{Mead:1979/jcp/2284}%
  \BibitemOpen
  \bibfield  {author} {\bibinfo {author} {\bibfnamefont {C.~A.}\ \bibnamefont
  {Mead}}\ and\ \bibinfo {author} {\bibfnamefont {D.~G.}\ \bibnamefont
  {Truhlar}},\ }\href {\doibase 10.1063/1.437734} {\bibfield  {journal}
  {\bibinfo  {journal} {J. Chem. Phys.}\ }\textbf {\bibinfo {volume} {70}},\
  \bibinfo {pages} {2284} (\bibinfo {year} {1979})}\BibitemShut {NoStop}%
\bibitem [{\citenamefont {Berry}(1987)}]{Berry:1987/rspa/31}%
  \BibitemOpen
  \bibfield  {author} {\bibinfo {author} {\bibfnamefont {M.~V.}\ \bibnamefont
  {Berry}},\ }\href {\doibase 10.1098/rspa.1987.0131} {\bibfield  {journal}
  {\bibinfo  {journal} {Proc. R. Soc. A}\ }\textbf {\bibinfo {volume} {414}},\
  \bibinfo {pages} {31} (\bibinfo {year} {1987})}\BibitemShut {NoStop}%
\bibitem [{\citenamefont {Sch\"{o}n}\ and\ \citenamefont
  {K\"{o}ppel}(1995)}]{Schon:1995/jcp/9292}%
  \BibitemOpen
  \bibfield  {author} {\bibinfo {author} {\bibfnamefont {J.}~\bibnamefont
  {Sch\"{o}n}}\ and\ \bibinfo {author} {\bibfnamefont {H.}~\bibnamefont
  {K\"{o}ppel}},\ }\href {\doibase 10.1063/1.469988} {\bibfield  {journal}
  {\bibinfo  {journal} {J. Chem. Phys.}\ }\textbf {\bibinfo {volume} {103}},\
  \bibinfo {pages} {9292} (\bibinfo {year} {1995})}\BibitemShut {NoStop}%
\bibitem [{\citenamefont {Juanes-Marcos}\ \emph {et~al.}(2005)\citenamefont
  {Juanes-Marcos}, \citenamefont {Althorpe},\ and\ \citenamefont
  {Wrede}}]{JuanesMarcos:2005/sci/1227}%
  \BibitemOpen
  \bibfield  {author} {\bibinfo {author} {\bibfnamefont {J.~C.}\ \bibnamefont
  {Juanes-Marcos}}, \bibinfo {author} {\bibfnamefont {S.~C.}\ \bibnamefont
  {Althorpe}}, \ and\ \bibinfo {author} {\bibfnamefont {E.}~\bibnamefont
  {Wrede}},\ }\href {\doibase 10.1126/science.1114890} {\bibfield  {journal}
  {\bibinfo  {journal} {Science}\ }\textbf {\bibinfo {volume} {309}},\ \bibinfo
  {pages} {1227} (\bibinfo {year} {2005})}\BibitemShut {NoStop}%
\bibitem [{\citenamefont {Kendrick}(2003)}]{Kendrick:2003}%
  \BibitemOpen
  \bibfield  {author} {\bibinfo {author} {\bibfnamefont {B.~K.}\ \bibnamefont
  {Kendrick}},\ }in\ \href {\doibase 10.1142/9789812565464_0012} {\emph
  {\bibinfo {booktitle} {{Conical Intersections. Electronic Structure, Dynamics
  and Spectroscopy}}}},\ \bibinfo {series} {Advanced Series in Physical
  Chemistry}, Vol.~\bibinfo {volume} {15},\ \bibinfo {editor} {edited by\
  \bibinfo {editor} {\bibfnamefont {W.}~\bibnamefont {Domcke}}, \bibinfo
  {editor} {\bibfnamefont {D.~R.}\ \bibnamefont {Yarkony}}, \ and\ \bibinfo
  {editor} {\bibfnamefont {H.}~\bibnamefont {K\"oppel}}}\ (\bibinfo
  {publisher} {World Scientific},\ \bibinfo {year} {2003})\ Chap.~\bibinfo
  {chapter} {12}, pp.\ \bibinfo {pages} {521--553}\BibitemShut {NoStop}%
\bibitem [{\citenamefont {Hazra}\ \emph {et~al.}(2015)\citenamefont {Hazra},
  \citenamefont {Balakrishnan},\ and\ \citenamefont {Kendrick}}]{Hazra:2015he}%
  \BibitemOpen
  \bibfield  {author} {\bibinfo {author} {\bibfnamefont {J.}~\bibnamefont
  {Hazra}}, \bibinfo {author} {\bibfnamefont {N.}~\bibnamefont {Balakrishnan}},
  \ and\ \bibinfo {author} {\bibfnamefont {B.~K.}\ \bibnamefont {Kendrick}},\
  }\href@noop {} {\bibfield  {journal} {\bibinfo  {journal} {Nat. Commun.}\
  }\textbf {\bibinfo {volume} {6}},\ \bibinfo {pages} {1} (\bibinfo {year}
  {2015})}\BibitemShut {NoStop}%
\bibitem [{\citenamefont {Althorpe}\ \emph {et~al.}(2008)\citenamefont
  {Althorpe}, \citenamefont {Stecher},\ and\ \citenamefont
  {Bouakline}}]{Althorpe:2008/jcp/214117}%
  \BibitemOpen
  \bibfield  {author} {\bibinfo {author} {\bibfnamefont {S.~C.}\ \bibnamefont
  {Althorpe}}, \bibinfo {author} {\bibfnamefont {T.}~\bibnamefont {Stecher}}, \
  and\ \bibinfo {author} {\bibfnamefont {F.}~\bibnamefont {Bouakline}},\ }\href
  {\doibase 10.1063/1.3031215} {\bibfield  {journal} {\bibinfo  {journal} {J.
  Chem. Phys.}\ }\textbf {\bibinfo {volume} {129}},\ \bibinfo {pages} {214117}
  (\bibinfo {year} {2008})}\BibitemShut {NoStop}%
\bibitem [{\citenamefont {Joubert-Doriol}\ \emph {et~al.}(2013)\citenamefont
  {Joubert-Doriol}, \citenamefont {Ryabinkin},\ and\ \citenamefont
  {Izmaylov}}]{Joubert:2013/jcp/234103}%
  \BibitemOpen
  \bibfield  {author} {\bibinfo {author} {\bibfnamefont {L.}~\bibnamefont
  {Joubert-Doriol}}, \bibinfo {author} {\bibfnamefont {I.~G.}\ \bibnamefont
  {Ryabinkin}}, \ and\ \bibinfo {author} {\bibfnamefont {A.~F.}\ \bibnamefont
  {Izmaylov}},\ }\href {\doibase http://dx.doi.org/10.1063/1.4844095}
  {\bibfield  {journal} {\bibinfo  {journal} {J. Chem. Phys.}\ }\textbf
  {\bibinfo {volume} {139}},\ \bibinfo {pages} {234103} (\bibinfo {year}
  {2013})}\BibitemShut {NoStop}%
\bibitem [{\citenamefont {Ryabinkin}\ \emph {et~al.}(2014)\citenamefont
  {Ryabinkin}, \citenamefont {Joubert-Doriol},\ and\ \citenamefont
  {Izmaylov}}]{Ryabinkin:2014/jcp/214116}%
  \BibitemOpen
  \bibfield  {author} {\bibinfo {author} {\bibfnamefont {I.~G.}\ \bibnamefont
  {Ryabinkin}}, \bibinfo {author} {\bibfnamefont {L.}~\bibnamefont
  {Joubert-Doriol}}, \ and\ \bibinfo {author} {\bibfnamefont {A.~F.}\
  \bibnamefont {Izmaylov}},\ }\href {\doibase
  http://dx.doi.org/10.1063/1.4881147} {\bibfield  {journal} {\bibinfo
  {journal} {J. Chem. Phys.}\ }\textbf {\bibinfo {volume} {140}},\ \bibinfo
  {pages} {214116} (\bibinfo {year} {2014})}\BibitemShut {NoStop}%
\bibitem [{\citenamefont {Bouakline}(2014)}]{Bouakline:2014/cp/31}%
  \BibitemOpen
  \bibfield  {author} {\bibinfo {author} {\bibfnamefont {F.}~\bibnamefont
  {Bouakline}},\ }\href {\doibase 10.1016/j.chemphys.2014.02.010} {\bibfield
  {journal} {\bibinfo  {journal} {Chemical Physics}\ }\textbf {\bibinfo
  {volume} {442}},\ \bibinfo {pages} {31} (\bibinfo {year} {2014})}\BibitemShut
  {NoStop}%
\bibitem [{\citenamefont {Xie}\ \emph {et~al.}(2016)\citenamefont {Xie},
  \citenamefont {Ma}, \citenamefont {Zhu}, \citenamefont {Yarkony},
  \citenamefont {Xie},\ and\ \citenamefont {Guo}}]{Xie:2016/jacs/7828}%
  \BibitemOpen
  \bibfield  {author} {\bibinfo {author} {\bibfnamefont {C.}~\bibnamefont
  {Xie}}, \bibinfo {author} {\bibfnamefont {J.}~\bibnamefont {Ma}}, \bibinfo
  {author} {\bibfnamefont {X.}~\bibnamefont {Zhu}}, \bibinfo {author}
  {\bibfnamefont {D.~R.}\ \bibnamefont {Yarkony}}, \bibinfo {author}
  {\bibfnamefont {D.}~\bibnamefont {Xie}}, \ and\ \bibinfo {author}
  {\bibfnamefont {H.}~\bibnamefont {Guo}},\ }\href {\doibase
  10.1021/jacs.6b03288} {\bibfield  {journal} {\bibinfo  {journal} {J. Am.
  Chem. Soc.}\ }\textbf {\bibinfo {volume} {138}},\ \bibinfo {pages} {7828}
  (\bibinfo {year} {2016})}\BibitemShut {NoStop}%
\bibitem [{\citenamefont {Gozem}\ \emph {et~al.}(2012)\citenamefont {Gozem},
  \citenamefont {Schapiro}, \citenamefont {Ferre},\ and\ \citenamefont
  {Olivucci}}]{Gozem:2012kg}%
  \BibitemOpen
  \bibfield  {author} {\bibinfo {author} {\bibfnamefont {S.}~\bibnamefont
  {Gozem}}, \bibinfo {author} {\bibfnamefont {I.}~\bibnamefont {Schapiro}},
  \bibinfo {author} {\bibfnamefont {N.}~\bibnamefont {Ferre}}, \ and\ \bibinfo
  {author} {\bibfnamefont {M.}~\bibnamefont {Olivucci}},\ }\href@noop {}
  {\bibfield  {journal} {\bibinfo  {journal} {Science}\ }\textbf {\bibinfo
  {volume} {337}},\ \bibinfo {pages} {1225} (\bibinfo {year}
  {2012})}\BibitemShut {NoStop}%
\bibitem [{\citenamefont {Tscherbul}\ and\ \citenamefont
  {Brumer}(2015)}]{Tscherbul:pccp/2015}%
  \BibitemOpen
  \bibfield  {author} {\bibinfo {author} {\bibfnamefont {T.~V.}\ \bibnamefont
  {Tscherbul}}\ and\ \bibinfo {author} {\bibfnamefont {P.}~\bibnamefont
  {Brumer}},\ }\href@noop {} {\bibfield  {journal} {\bibinfo  {journal} {Phys.
  Chem. Chem. Phys.}\ }\textbf {\bibinfo {volume} {17}},\ \bibinfo {pages}
  {30904} (\bibinfo {year} {2015})}\BibitemShut {NoStop}%
\bibitem [{\citenamefont {K\"{o}ppel}\ \emph {et~al.}(1984)\citenamefont
  {K\"{o}ppel}, \citenamefont {Domcke},\ and\ \citenamefont
  {Cederbaum}}]{Koppel:1984/acp/59}%
  \BibitemOpen
  \bibfield  {author} {\bibinfo {author} {\bibfnamefont {H.}~\bibnamefont
  {K\"{o}ppel}}, \bibinfo {author} {\bibfnamefont {W.}~\bibnamefont {Domcke}},
  \ and\ \bibinfo {author} {\bibfnamefont {L.~S.}\ \bibnamefont {Cederbaum}},\
  }\enquote {\bibinfo {title} {{Multimode Molecular Dynamics Beyond the
  Born-Oppenheimer Approximation}},}\ \ (\bibinfo  {publisher} {John Wiley \&
  Sons, Inc.},\ \bibinfo {year} {1984})\ Chap.~\bibinfo {chapter} {2}, pp.\
  \bibinfo {pages} {59--246}\BibitemShut {NoStop}%
\bibitem [{\citenamefont {Mead}(1982)}]{Mead:1982/jcp/6090}%
  \BibitemOpen
  \bibfield  {author} {\bibinfo {author} {\bibfnamefont {C.~A.}\ \bibnamefont
  {Mead}},\ }\href {\doibase 10.1063/1.443853} {\bibfield  {journal} {\bibinfo
  {journal} {J. Chem. Phys.}\ }\textbf {\bibinfo {volume} {77}},\ \bibinfo
  {pages} {6090} (\bibinfo {year} {1982})}\BibitemShut {NoStop}%
\bibitem [{\citenamefont {Cederbaum}\ \emph {et~al.}(1977)\citenamefont
  {Cederbaum}, \citenamefont {Domcke}, \citenamefont {K{\"{o}}ppel},\ and\
  \citenamefont {Von~Niessen}}]{Cederbaum:1977/cp/169}%
  \BibitemOpen
  \bibfield  {author} {\bibinfo {author} {\bibfnamefont {L.}~\bibnamefont
  {Cederbaum}}, \bibinfo {author} {\bibfnamefont {W.}~\bibnamefont {Domcke}},
  \bibinfo {author} {\bibfnamefont {H.}~\bibnamefont {K{\"{o}}ppel}}, \ and\
  \bibinfo {author} {\bibfnamefont {W.}~\bibnamefont {Von~Niessen}},\ }\href
  {\doibase 10.1016/0301-0104(77)87041-9} {\bibfield  {journal} {\bibinfo
  {journal} {Chem. Phys}\ }\textbf {\bibinfo {volume} {26}},\ \bibinfo {pages}
  {169} (\bibinfo {year} {1977})}\BibitemShut {NoStop}%
\bibitem [{\citenamefont {Sukharev}\ and\ \citenamefont
  {Seideman}(2005)}]{Sukharev:2005/pra/012509}%
  \BibitemOpen
  \bibfield  {author} {\bibinfo {author} {\bibfnamefont {M.}~\bibnamefont
  {Sukharev}}\ and\ \bibinfo {author} {\bibfnamefont {T.}~\bibnamefont
  {Seideman}},\ }\href {\doibase 10.1103/PhysRevA.71.012509} {\bibfield
  {journal} {\bibinfo  {journal} {Phys. Rev. A}\ }\textbf {\bibinfo {volume}
  {71}},\ \bibinfo {pages} {012509} (\bibinfo {year} {2005})}\BibitemShut
  {NoStop}%
\bibitem [{\citenamefont {Gindensperger}\ \emph
  {et~al.}(2006{\natexlab{a}})\citenamefont {Gindensperger}, \citenamefont
  {Burghardt},\ and\ \citenamefont
  {Cederbaum}}]{Gindensperger:2006/jcp/144104}%
  \BibitemOpen
  \bibfield  {author} {\bibinfo {author} {\bibfnamefont {E.}~\bibnamefont
  {Gindensperger}}, \bibinfo {author} {\bibfnamefont {I.}~\bibnamefont
  {Burghardt}}, \ and\ \bibinfo {author} {\bibfnamefont {L.~S.}\ \bibnamefont
  {Cederbaum}},\ }\href {\doibase 10.1063/1.2183305} {\bibfield  {journal}
  {\bibinfo  {journal} {J. Chem. Phys.}\ }\textbf {\bibinfo {volume} {124}},\
  \bibinfo {pages} {144104} (\bibinfo {year} {2006}{\natexlab{a}})}\BibitemShut
  {NoStop}%
\bibitem [{\citenamefont {Izmaylov}\ \emph {et~al.}(2011)\citenamefont
  {Izmaylov}, \citenamefont {{Mendive-Tapia}}, \citenamefont {Bearpark},
  \citenamefont {Robb}, \citenamefont {Tully},\ and\ \citenamefont
  {Frisch}}]{Izmaylov:2011/jcp/234106}%
  \BibitemOpen
  \bibfield  {author} {\bibinfo {author} {\bibfnamefont {A.~F.}\ \bibnamefont
  {Izmaylov}}, \bibinfo {author} {\bibfnamefont {D.}~\bibnamefont
  {{Mendive-Tapia}}}, \bibinfo {author} {\bibfnamefont {M.~J.}\ \bibnamefont
  {Bearpark}}, \bibinfo {author} {\bibfnamefont {M.~A.}\ \bibnamefont {Robb}},
  \bibinfo {author} {\bibfnamefont {J.~C.}\ \bibnamefont {Tully}}, \ and\
  \bibinfo {author} {\bibfnamefont {M.~J.}\ \bibnamefont {Frisch}},\ }\href
  {\doibase 10.1063/1.3667203} {\bibfield  {journal} {\bibinfo  {journal} {J.
  Chem. Phys.}\ }\textbf {\bibinfo {volume} {135}},\ \bibinfo {pages} {234106}
  (\bibinfo {year} {2011})}\BibitemShut {NoStop}%
\bibitem [{\citenamefont {Thiel}\ and\ \citenamefont
  {Köppel}(1999)}]{Thiel:1999/jcp/9372}%
  \BibitemOpen
  \bibfield  {author} {\bibinfo {author} {\bibfnamefont {A.}~\bibnamefont
  {Thiel}}\ and\ \bibinfo {author} {\bibfnamefont {H.}~\bibnamefont
  {Köppel}},\ }\href {\doibase 10.1063/1.478902} {\bibfield  {journal}
  {\bibinfo  {journal} {J. Chem. Phys.}\ }\textbf {\bibinfo {volume} {110}},\
  \bibinfo {pages} {9371} (\bibinfo {year} {1999})}\BibitemShut {NoStop}%
\bibitem [{\citenamefont {K\"{o}ppel}\ \emph {et~al.}(2001)\citenamefont
  {K\"{o}ppel}, \citenamefont {Gronki},\ and\ \citenamefont
  {Mahapatra}}]{Koppel:2001/jcp/2377}%
  \BibitemOpen
  \bibfield  {author} {\bibinfo {author} {\bibfnamefont {H.}~\bibnamefont
  {K\"{o}ppel}}, \bibinfo {author} {\bibfnamefont {J.}~\bibnamefont {Gronki}},
  \ and\ \bibinfo {author} {\bibfnamefont {S.}~\bibnamefont {Mahapatra}},\
  }\href {\doibase 10.1063/1.1383986} {\bibfield  {journal} {\bibinfo
  {journal} {J. Chem. Phys.}\ }\textbf {\bibinfo {volume} {115}},\ \bibinfo
  {pages} {2377} (\bibinfo {year} {2001})}\BibitemShut {NoStop}%
\bibitem [{\citenamefont {Cederbaum}\ \emph {et~al.}(2005)\citenamefont
  {Cederbaum}, \citenamefont {Gindensperger},\ and\ \citenamefont
  {Burghardt}}]{Cederbaum:2005/prl/113003}%
  \BibitemOpen
  \bibfield  {author} {\bibinfo {author} {\bibfnamefont {L.}~\bibnamefont
  {Cederbaum}}, \bibinfo {author} {\bibfnamefont {E.}~\bibnamefont
  {Gindensperger}}, \ and\ \bibinfo {author} {\bibfnamefont {I.}~\bibnamefont
  {Burghardt}},\ }\href {\doibase 10.1103/PhysRevLett.94.113003} {\bibfield
  {journal} {\bibinfo  {journal} {Phys. Rev. Lett.}\ }\textbf {\bibinfo
  {volume} {94}},\ \bibinfo {pages} {113003} (\bibinfo {year}
  {2005})}\BibitemShut {NoStop}%
\bibitem [{\citenamefont {Gindensperger}\ \emph
  {et~al.}(2006{\natexlab{b}})\citenamefont {Gindensperger}, \citenamefont
  {Burghardt},\ and\ \citenamefont
  {Cederbaum}}]{Gindensperger:2006/jcp/144103}%
  \BibitemOpen
  \bibfield  {author} {\bibinfo {author} {\bibfnamefont {E.}~\bibnamefont
  {Gindensperger}}, \bibinfo {author} {\bibfnamefont {I.}~\bibnamefont
  {Burghardt}}, \ and\ \bibinfo {author} {\bibfnamefont {L.~S.}\ \bibnamefont
  {Cederbaum}},\ }\href {\doibase 10.1063/1.2183304} {\bibfield  {journal}
  {\bibinfo  {journal} {J. Chem. Phys.}\ }\textbf {\bibinfo {volume} {124}},\
  \bibinfo {pages} {144103} (\bibinfo {year} {2006}{\natexlab{b}})}\BibitemShut
  {NoStop}%
\bibitem [{\citenamefont {Gherib}\ \emph {et~al.}(2016)\citenamefont {Gherib},
  \citenamefont {Ye}, \citenamefont {Ryabinkin},\ and\ \citenamefont
  {Izmaylov}}]{Gherib:2016ch}%
  \BibitemOpen
  \bibfield  {author} {\bibinfo {author} {\bibfnamefont {R.}~\bibnamefont
  {Gherib}}, \bibinfo {author} {\bibfnamefont {L.}~\bibnamefont {Ye}}, \bibinfo
  {author} {\bibfnamefont {I.~G.}\ \bibnamefont {Ryabinkin}}, \ and\ \bibinfo
  {author} {\bibfnamefont {A.~F.}\ \bibnamefont {Izmaylov}},\ }\href@noop {}
  {\bibfield  {journal} {\bibinfo  {journal} {J. Chem. Phys.}\ }\textbf
  {\bibinfo {volume} {144}},\ \bibinfo {pages} {154103} (\bibinfo {year}
  {2016})}\BibitemShut {NoStop}%
\bibitem [{\citenamefont {Gherib}\ \emph {et~al.}(2015)\citenamefont {Gherib},
  \citenamefont {Ryabinkin},\ and\ \citenamefont
  {Izmaylov}}]{Gherib:2015/jctc/11}%
  \BibitemOpen
  \bibfield  {author} {\bibinfo {author} {\bibfnamefont {R.}~\bibnamefont
  {Gherib}}, \bibinfo {author} {\bibfnamefont {I.~G.}\ \bibnamefont
  {Ryabinkin}}, \ and\ \bibinfo {author} {\bibfnamefont {A.~F.}\ \bibnamefont
  {Izmaylov}},\ }\href@noop {} {\bibfield  {journal} {\bibinfo  {journal} {J.
  Chem. Theory and Comput.}\ }\textbf {\bibinfo {volume} {11}},\ \bibinfo
  {pages} {1375} (\bibinfo {year} {2015})}\BibitemShut {NoStop}%
\bibitem [{\citenamefont {Tannor}(2007)}]{Tannor:2007}%
  \BibitemOpen
  \bibfield  {author} {\bibinfo {author} {\bibfnamefont {D.~J.}\ \bibnamefont
  {Tannor}},\ }\href@noop {} {\emph {\bibinfo {title} {Introduction to Quantum
  Dynamics: A Time-Dependent Perspective}}}\ (\bibinfo  {publisher} {University
  Science Books},\ \bibinfo {address} {Sausalito, Cal},\ \bibinfo {year}
  {2007})\BibitemShut {NoStop}%
\end{thebibliography}
%
\end{document}